\documentclass[%
 reprint,
superscriptaddress,
nofootinbib,
 amsmath,amssymb,
 aps,
floatfix
]{revtex4-2}

\usepackage{graphicx}% Include figure files
\usepackage{dcolumn}% Align table columns on decimal point
\usepackage{bm}% bold math
\usepackage{aas_macros}
\usepackage{xcolor} 

\definecolor{green}{rgb}{0,0.4,0}

\usepackage{hyperref}
\hypersetup{linkcolor=red,citecolor=blue,urlcolor=cyan}

\begin{document}

\title{Differentiating between sharp and smoother phase transitions in neutron stars} 

\author{Jonas P. Pereira}%
\email{jpereira@camk.edu.pl}
\affiliation{Nicolaus Copernicus Astronomical Center, Polish Academy of Sciences, Bartycka 18, 00-716, Warsaw, Poland}

\author{Micha{\l} Bejger}
%\email{bejger@camk.edu.pl}
\affiliation{INFN Sezione di Ferrara, Via Saragat 1, 44122 Ferrara, Italy}
\affiliation{Nicolaus Copernicus Astronomical Center, Polish Academy of Sciences, Bartycka 18, 00-716, Warsaw, Poland}

\author{J. Leszek Zdunik}%
%\email{jlz@camk.edu.pl}
\affiliation{Nicolaus Copernicus Astronomical Center, Polish Academy of Sciences, Bartycka 18, 00-716, Warsaw, Poland}

\author{Pawe{\l} Haensel}%
%\email{haensel@camk.edu.pl}
\affiliation{Nicolaus Copernicus Astronomical Center, Polish Academy of Sciences, Bartycka 18, 00-716, Warsaw, Poland}

\date{\today}% It is always \today, today,

\begin{abstract}
The internal composition of neutron stars is still an open issue in astrophysics. Their innermost regions are impervious to light propagation and gravitational waves mostly carry global aspects of stars, meaning that only indirect inferences of their interiors could be obtained. Here we  assume a hypothetical future scenario in which an equation of state softening due to a phase transition is identified and estimate the observational accuracy to differentiate 
%mixed phase/state from 
a sharp phase transition from a smoother one (which we take to be associated with a mixed phase/state due to the unknown value of the surface tension of dense matter) in  a region of a hybrid star by means of some electromagnetic and gravitational wave observables. We show that different transition constructions lead to similar sequences of stellar configurations due to their shared thermodynamic properties. In the most optimistic case - a strong quark-hadron density jump phase transition - radius observations require fractional uncertainties smaller than $1\%-2\%$ to differentiate mixed states from sharp phase transitions. For tidal deformabilities, relative uncertainties should be smaller than $5\%-10\%$. However, for masses around the onset of stable quark cores, relative tidal deformability differences associated with strong sharp phase transitions and mixed states connecting the two pure phases could be much larger (up to around $20\%-30\%$). 
All the above suggests that 2.5- and 3rd generation gravitational wave detectors and near-term electromagnetic missions may be able to start assessing some particular aspects of phase transitions in neutron stars. In addition, it points to some limitations on the equation of state recovery using typical neutron star observables and the impact of systematic uncertainties on modellings of the equation of state of hybrid stars.
Finally, we briefly discuss other observables that may also be relevant for the probe of mixed states in stars.
%\begin{description}
%... if needed 
%\end{description}
\end{abstract}

%\keywords{Suggested keywords}%
\maketitle

%\tableofcontents

\section{Introduction}
\label{sect:intro} 
It has long been hypothesized that neutron stars (NSs) contain exotic phases of matter, which are possible to exist solely due to unique conditions - density and pressure, particle fractions - present in their interiors \cite{HaenselPY2007}. The change from ``normal'' (nucleonic) matter to ``exotic'' matter (e.g. deconfined quarks) is thought to occur through a phase transition (or, in more general terms, state transition) process. Its detailed structure - sharp first order between the two pure phases or via a mixed state of the two phases - is still an open issue \citep{2011RPPh...74a4001F,2017EPJA...53...60A}, due to the complexity of direct quantum chromodynamics (QCD) calculations \citep{2018RPPh...81h4301R}, or the lack of direct experimental observations of dense matter at high chemical potentials and low temperatures. The above clearly shows that the most promising laboratories for studying dense matter aspects are NSs \citep{2008RvMP...80.1455A,2016EPJA...52...49F,2018RPPh...81e6902B,2019JPhG...46g3002O}. In general, a state transition between distinct phases of matter results in a {\em softening} of the pressure-density relation in the equation of state (EOS), which in turn results in more compact NSs (in terms of stellar parameters, this is quantified by larger values of the compactness parameter $GM/Rc^2$, with $M$ denoting the gravitational mass, and $R$ the stellar radius), and a lower maximum mass $M_{\rm max}$ than in the case of stars without state transitions, due to transitional deficit in pressure increase related to the softening. While a direct access to the interiors of NS is impossible, one can draw conclusions from astrophysical measurements of the stellar mass $M$ and radius $R$ with the use of electromagnetic observables (see, e.g., \cite{2019ApJ...887L..24M,2019ApJ...887L..21R,2021ApJ...918L..28M,2021arXiv210506980R,2021arXiv210506981R}), as well as the tidal deformabilities $\Lambda_i$ of the components of a binary system during its last orbits before the merger by means of gravitational wave (GW) signals (\cite{2008PhRvD..77b1502F,vanOeveren2017,Abbott2017a,Abbott2017b,2020PhRvD.101l3007L,2020PhRvD.101f3007E,2020PhRvR...2c3514P}, see also \cite{2020GReGr..52..109C} for a review), due to their dependence on the EOS; hence, one can expect potentially measurable imprints of dense-matter state transitions on these NS observables. For recent reviews on the dense-matter state transitions in NSs, see, e.g., \cite{2019arXiv190103252D,PhysRevD.101.044019}. 

Here we study a hypothetical future scenario in which a dense-matter softening is identified by means of global NS parameters, e.g. $M$, $R$ and $\Lambda$. We model it by assuming that a {\em detailed} behavior of the microscopic quantities in the EOS, in the range of EOS where the softening occurs is {\em de facto} unknown, and - for simplicity - assume that the transition from a nucleonic outer part of an NS to an inner (possibly exotic) core happens via either a sharp boundary between pure phases (``density jump'' phase transition), or via a transition through a ``mixed phases'' (mixed state) region.

In addition we assume that the low-density part of the EOS (a ``nucleonic matter crust and outer core'') and the high-density part of the EOS (an ``exotic inner core'') are the same for given realisations of sharp phase transitions and mixed-state transitions that will be compared, i.e., we will study the state transition {\it masquerade problem}, similar to the one first discussed in \cite{2005ApJ...629..969A}. This setup realistically captures, in our opinion, the crucial observational difficulty we will encounter in the future. A key observation in this regard is a demonstration that the sequences of $M$, $R$, $\Lambda$ labeled by central stellar parameters (e.g., central pressure or central energy density) {\em outside} the range of values corresponding to the state-transition region depend very weakly on the exact details of the EOS behavior {\em within} the state-transition region, a feature demonstrated in Sect.~\ref{sect:equality_of_mmax} and in the exemplary Fig.~\ref{fig:mixed4}. All the above also suggests possible limitations for the EOS recovery with the use of common observables and characterizing on what level, in terms of accuracy, such a recovery should happen is a relevant issue.

A softening of the EOS could also be related to higher order phase transitions. Recent models of this type can be  found  in \cite{2019ApJ...885...42B,2018RPPh...81e6902B}, who consider  a  crossover  type transition between  hadronic and quark matter with a continuous speed of sound. The softening in the EOS, obtained in the quarkyonic matter model for the transition between hadrons and quarks,  keeps continuous the second derivative of the speed of sound \cite{2018qcs..confa1010X}. Also, there are other transitions to exotic states of hadronic matter involving a condensation of pions \cite{1973JETPL..18..260M,1973JETPL..18..260M,1972PhRvL..29..382S,1973PhRvD...7..953S} or kaons \cite{1986PhLB..175...57K}. Depending on the choice of parameters for the strong interaction Lagrangian, the softening of the EOS due to the boson condensed state  may be associated with a second order phase transition  or a first order one where a mixed state can be contemplated. Here, however, we interpret the softening of the EOS as due to a mixed state of hadronic and quark matter (or a sharp phase transition) because it is directly connected with the surface tension at the phases' interface \citep{2018PhRvC..97d5802A}.

The main motivation of this work is to study phase-transition parametric EOSs together with their corresponding sequences of NS parameters, in order to establish {\em how much} the {\em a priori} unknown state-transition imprints on the global NS parameters ($M$, $R$, $\Lambda$), and to find possible characteristic features of either sharp or a mixed-state transitions in these observables. Specifically, we assess the critical accuracy of the current and planned observing infrastructures, necessary to falsify specific dense-matter models. 
In order to come up with optimistic upper limits for the accuracies, we focus on the most extreme phase transition case (strong phase transition \citep{2019A&A...622A.174S}), directly related to the existence of the third family of stars \citep{2017PhRvC..96d5809A,2017PhRvL.119p1104A,2018PhRvC..97d5802A,2019arXiv191209809C,2019PhRvC.100b5802M}. This is so because it would imply that NSs with similar masses could have very different radii, which is a direct result of a destabilization of stellar configurations for a range of central pressures larger than the one marking a state transition. 
(``Weak'' phase transitions would produce smaller differences in the NS parameters.)

In the near future it will be possible to constrain masses and radii of NS (see, e.g., \citep{2016SPIE.9905E..1QZ,2019SCPMA..6229503W}) and tidal deformabilities with uncertainties of a few percent, the latter with the third generation GW detectors events with high signal-to-noise ratios (see, e.g., \citep{2021arXiv210812368C} and references therein), with a real possibility of assessing the nature of an observed soft interval in the
EoS.
While we do not directly focus on stability issues associated with the mixed state. For an analysis in this direction, see \citep{2018PhRvC..97d5802A}. Stability should be easily identified from the $M(R)$ sequences with the configurations to the right (left) of their maxima (minima); in general, $\partial M/\partial \rho_c\geq 0$, where $\rho_c$ is the central density, see, e.g., \citep{2018ApJ...860...12P} and references therein. Studies on the stability of rotating hybrid stars with mixed states and sharp phase transitions indicate that rotation does not change the global property of the (non)existence of the second branch of stable configurations \citep{2006A&A...450..747Z}.

Given our partial ignorance of the EOS around and above the nuclear saturation density, we explore here many models by means of parameterizations of the EOS (for the high-density part and also for the mixed state). We vary the parameters in a way to cover our expectations (for strong phase transitions and mixed states) and some available NS constraints. Clearly, this model is phenomenological, and could encompass many microscopic descriptions for the softening of the EOS by means of different parameter choices. We draw conclusions about the accuracies needed for distinguishing a sharp phase transition from a mixed state from these analyses. 
Naturally, this study is not exhaustive but rather indicative of the relevant cases to better focus on more precise future works.

The article is composed as follows: in Sect.~\ref{sect:eos} we describe two parametric models of the EOSs (model of \citeauthor{2018Univ....4...94A} \citeyear{2018Univ....4...94A} \citep{2018Univ....4...94A} in Sect.~\ref{sect:blaschke}, and a piecewise-polytropic model in Sect.~\ref{sect:polytropes}), which will be used in Sect.~\ref{sect:tov} as an input to the Tolman-Oppenheimer-Volkoff (TOV) equations \cite{1939PhRv...55..364T,1939PhRv...55..374O} to produce sequences of $M$, $R$ and $\Lambda$ as functions of central pressure $P$ and chemical potential $\mu$. Specifically, Sect.~\ref{sect:results} discusses how the EOS difference between the ``sharp'' or ``mixed'' state transitions regions impacts the $M(R)$ and $M(\Lambda)$ sequences, as well as the value of the maximum mass $M_{\rm max}$, and assess the regions of astrophysical parameters, for which the ``sharp'' and ``mixed'' state transitions result in potentially observable differences (a state-transition {\it masquerade problem}). Section~\ref{sect:results} contains also a discussion of these results from the point of view of a current and planned capabilities of the EM missions (NICER \cite{NICER}, Athena \cite{2013arXiv1306.2307N}, eXTP \cite{2016SPIE.9905E..1QZ,2019SCPMA..6229503W}) and the GW detectors (Advanced LIGO \cite{2015CQGra..32g4001L}, Advanced Virgo \cite{2015CQGra..32b4001A}, KAGRA \cite{2019NatAs...3...35K}, NEMO \citep{2020PASA...37...47A}, Einstein Telescope \cite{2020JCAP...03..050M}, Cosmic Explorer \cite{2019BAAS...51g..35R}) in terms of measurements errors. Section~\ref{sect:conclusions} contains a relevant discussion, conclusions and an outlook. Section~\ref{summary} gives a detailed summary of our analysis.

\section{Parametric models of the EOS} 
\label{sect:eos} 

The mixed phase/state in a hybrid star can be approximated in a variety of ways. 
Under the theoretical viewpoint, a macroscopically smoother phase transition (leading to the presence of an intermediate state--mixed phase/state) in a star---besides a sharp phase transition---is thermodynamically possible. Indeed, pasta phases \citep{2001PhR...342..393G,2018ASSL..457..337B} for dense matter, where nuclei exist in non-spherical shapes, could be present in the bottom layer of an NS crust. In addition, there is plenty of room in models for hybrid stars where third families of NSs \citep{2017PhRvC..96d5809A,2017PhRvL.119p1104A,2018PhRvC..97d5802A,2019arXiv191209809C,2019PhRvC.100b5802M} are very distinct from purely hadronic stars, meaning that it would make observational sense to contrast a sharp transition with one having an intermediate state.

Here, in Sect.~\ref{sect:blaschke}, we present the construction put forth by \cite{2018Univ....4...94A} (for further applications, see \cite{2020Univ....6...81B}). Their main idea is to build on the Maxwell's construction (sharp phase transition) and phenomenologically take into account certain properties of other (microscopic) mixed state constructions. In Sect.~\ref{sect:polytropes} we present a simple sharp/mixed state transition based on the use of relativistic polytropes (\cite{Tooper1965}).

\subsection{Mixed state construction of~\citeauthor{2018Univ....4...94A}~\citeyear{2018Univ....4...94A}}
\label{sect:blaschke} 

Here we quickly review the mixed state construction as put forward by \citep{2018Univ....4...94A}. We denote by $\mu_{0}$ the baryon chemical potential at the quark-hadron phase transition coming from the Maxwell construction. One expects the presence of a mixed state to increase the phase transition pressure $P$ at $\mu_0$, $P_{0}\equiv P(\mu_{0})$. The reason for that is the larger density of the mixed state with respect to the base of the hadron phase. The relative increase of pressure may be assumed to be given, as motivated by first principle constructions, and will be denoted by $\Delta_p$. For the connection of $\Delta_p$ with microscopic parameters of the pasta phase, see \cite{2019PhRvC.100b5802M}. In addition, assume that the pressure in the mixed state, $P_m$, in the simplest case is given by (parabolic expansion)
\begin{equation}
    P_m(\mu) = (1+\Delta_p)P_{0} + \alpha_1(\mu-\mu_{0}) + \alpha_2(\mu-\mu_{0})^2\label{p_mixed},
\end{equation}
where $\alpha_1$ and $\alpha_2$ are free parameters to be found by demanding certain thermodynamic constraints. In particular, we impose the continuity of the mixed state pressure and its first derivative with respect to the baryon chemical potential (baryon number) at the hadronic ($\mu=\mu_h$) and quark ($\mu=\mu_q$) interfaces:
\begin{equation}
P_m(\mu_h)=P_h(\mu_h),\;\;\; P_m(\mu_q)=P_q(\mu_q)\label{p_condition}
\end{equation}
and
\begin{equation}
n_m(\mu_h)=n_h(\mu_h),\;\;\; n_m(\mu_q)=n_q(\mu_q)\label{n_condition},
\end{equation}
with $\mu_h$ and $\mu_q$ free adjustable quantities, $P_h$ the hadronic EOS and $P_q$ the quark EOS. Put in the above way, given a $\Delta_p$, one has a system of four equations (\ref{p_condition}, \ref{n_condition}) to four unknowns ($\mu_h,\mu_q,\alpha_1,\alpha_2$) to solve, and its solution should be unique. Obviously, the physically relevant solution should present $\mu_h<\mu_{0}<\mu_q$. After solving the TOV equations for a given central density, one can find the extension of the mixed state by means of the knowledge of $\mu_H$ and $\mu_Q$. In addition, the continuity of the baryon number density at both borders of the mixed state implies that the energy density is also continuous there for hadronic and quark barotropic EOSs. 

The speed of sound, $c_s$, on the other hand, is in general discontinuous at the hadronic and quark borders for the model given by Eq. \eqref{p_mixed}. The reason is simply because it involves a second derivative of the pressure (with respect to $\mu$), which is not guaranteed to be continuous at the borders of the mixed state for the parabolic mixed state construction. With the above prescription, it is not controllable and causality should be checked a posteriori. Given the causality of the speed of sound for both hadronic and quark phases and the expected EOS softening due to the mixed state, one would expect $c_s^2$ to also be causal there.

\subsection{Mixed state polytropic EOSs}
\label{sect:polytropes} 

Here we put forth an effective, parametric multi-polytrope model for both the sharp and mixed state transitions. Basic intensive thermodynamic properties of relativistic polytropes \cite{Tooper1965} are defined as:
\begin{eqnarray}
\label{eq:eospoly}
P(n)&=&Kn^\gamma,\nonumber\\
\rho(n)&=& n\varepsilon +\frac{P}{\gamma-1},\\
\mu(P)&=&\varepsilon + \frac{\gamma}{\gamma-1}\frac{P}{n},\nonumber
\end{eqnarray}
where $P$ is the pressure, $\rho$ the mass-energy density, $\mu$ the chemical potential, 
$\varepsilon$ the energy per baryon at $P=0$ in a given phase, $K$ is the polytropic ``pressure'' coefficient, and $\gamma$ is called the adiabatic index. Pressure and energy-density are functions of the baryon number $n$, but later we will focus on the direct relation between the chemical potential and the pressure, $\mu(P)$.\footnote{The $\gamma=1$ case needs a separate treatment (logarithmic dependence of $\rho(n)/n$) and is not considered here.} A sharp phase transition may be defined as a ``Maxwell construction'' at the first order phase transition point ($P_0$, $n_0$) between two polytropes ($K_1$, $\gamma_1$, $\varepsilon_1$) and ($K_2$, $\gamma_2$, $\varepsilon_2$), accompanied by the baryon number density jump $n_0=n_{01}\to n_{02}$, by the following condition resulting from the mechanical and chemical equilibrium of the associated phases: 
\begin{equation}
\overline \varepsilon_1 + \frac{\gamma_1}{\gamma_1-1}=\overline \varepsilon_2 + \frac{\gamma_2}{\gamma_2-1}\frac{1}{\lambda},
\label{eq:poly1poly2_m}
\end{equation}
where
\begin{equation}
\overline \varepsilon_i=\frac{\varepsilon_i n_0}{P_0}\quad \text{and}\quad \lambda=\frac{n_{02}}{n_{01}}=\frac{n_{02}}{n_0}.
\label{eq:poly1poly2_lambda}
\end{equation}

For the mixed state we assume a polytropic EOS given by Eq. \eqref{eq:eospoly} with parameters ($K_m$, $\gamma_m$, $\varepsilon_m$). Assuming the appearance of the mixed phase at a pressure $P_1<P_0$ (and the baryon density $n_1<n_0$), the parameters $\gamma_m$ and $\varepsilon_m$ of the ``mixed state'' polytrope are given by the solutions to the following relations: 
\begin{multline}
\left(\frac{1}{\gamma_2-1}-\frac{1}{\gamma_m-1}\right)\lambda^{\gamma_2(\gamma_m-1)/(\gamma_2-\gamma_m)}
\times \\ 
\overline{n}_1^{(\gamma_2-1)(\gamma_1-\gamma_m)/(\gamma_2-\gamma_m)} + \frac{\gamma_m}{\gamma_m-1}\overline n_1^{\gamma_1-1}= \\
\frac{\gamma_2}{(\gamma_2-1)}\frac{1}{\lambda}-\frac{\gamma_1}{\gamma_1-1}\left(1-\overline n_1^{\gamma_1-1}\right),
\label{eq:gammam}
\end{multline}
where $\overline{n}_1=n_1/n_0 <1$. The mean ``mixed'' value of the parameter $\varepsilon$ results from 
\begin{equation}
\overline{\varepsilon}_m=\overline{\varepsilon}_1+\left(\frac{\gamma_1}{\gamma_1-1}-\frac{\gamma_m}{\gamma_m-1}\right)
\overline{n}_1^{\gamma_1-1},
\label{eq:mm}
\end{equation}
whereas the endpoint of the mixed state ($P_3$, $n_3$) is determined by  
\begin{equation}
\frac{\overline P_3}{\overline n_3}=\frac{\overline n_3^{\gamma_2-1}}{\lambda^{\gamma_2}}=\frac{\overline \varepsilon_m-\overline \varepsilon_2}{\gamma_2/(\gamma_2-1)-\gamma_m/(\gamma_m-1)}.
\label{eq:p2n2}
\end{equation}
where ${\overline P_3}=P_3/P_0$.

Relativistic polytropes are used to define the dense ($n>n_{cc}$, where the subscript $cc$ denote the crust-core transition) part of the EOS. For the low-density part (the crust) we use the \citeauthor{DouchinH2001} SLy4 EOS \cite{DouchinH2001}. The SLy4 crust extends up to the pressure $P_{cc}$, densities $n_{cc}$, $\rho_{cc}$ and chemical potential $\mu_{cc}$. At $P=P_{cc}$, we define a smooth crust/core transition to the polytrope $P=K_1n^{\gamma_1}$ with one free parameter $\gamma_1$, and the other two parameters $(K_1$, $m_1)$ defined by:
\begin{equation}
K_1=\frac{P_{cc}}{n_{cc}^{\gamma_1}}\quad\text{and}\quad \varepsilon_1=\mu_{cc}-\frac{\gamma_1}{\gamma_1-1}. 
\end{equation}
At ($P_0$, $n_0$), a first-order phase transition between the polytropes ($K_1$, $\gamma_1$, $\varepsilon_1$) and ($K_2$, $\gamma_2$, $\varepsilon_2$) ( ``Maxwell construction'') is defined by the parameters in Eqs.~\eqref{eq:poly1poly2_m} and \eqref{eq:poly1poly2_lambda}. A polytrope with selected $\gamma_2$ and ($K_2$, $m_2$) resulting from the equilibrium conditions is given by 
\begin{equation}
K_2=\frac{P_0}{(\lambda n_0)^{\gamma_2}}\quad\text{and}\quad \overline \varepsilon_2 =\overline \varepsilon_1 +\frac{\gamma_1}{\gamma_1-1}-\frac{\gamma_2}{\gamma_2-1}\frac{1}{\lambda}.
\label{eq:poly2} 
\end{equation}

The mixed state is defined between $P_1$ and $P_3$, associated with the baryon numbers $n_1$ and $n_3$, respectively, with the point $P_1$($n_1$) being a free quantity to choose. With Eqs.~\eqref{eq:gammam}-\eqref{eq:p2n2} one obtains the parameters of the mixed-phase polytrope and the point ($P_3$, $n_3$). In general, for the mixed state, 
\begin{equation}
P_m=P_1\left(\frac{n}{n_1}\right)^{\gamma_m}\quad\text{and}\quad K_m=P_1/n_1^{\gamma_m}. 
\end{equation}
Note that the model described in Sect.~\ref{sect:blaschke} in the mixed region is also a specific case of a polytrope ($\gamma_m=2$) with an additional pressure term, equal to $(1+\Delta_p)P_{0}-\alpha^2_1/(2\alpha_2)$. In addition, this polytropic model is physically different from a hadron-quark crossover \citep{2013PTEP.2013g3D01M} because the speed of sound is not continuous on both borders of the mixed state in general. Although one would expect that to have a small impact on global stellar parameters, it might be important for other observables, such as oscillation modes. Also, the crossover nature of an EOS might even lead to its stiffening \citep{2013PTEP.2013g3D01M}, not generally the case for our current construction. We plan on extending our mixed state polytropic model to cover such possibilities in future works.

\section{Tidal deformability}
\label{sect:tov} 

The $M(R)$ sequences associated with the EOSs of Sec.\ref{sect:eos} come from the solution of the TOV system of equations assuming spherically symmetric spacetimes (see, e.g., \citep{HaenselPY2007}). For each background configuration, we have also calculated its tidal deformability. We have assumed perfect-fluids all along. Although some phases of NSs--especially the mixed states--should be elastic, we do not take into account this fact in this first study. The equation that we solve related to tidal deformations is \citep{2008ApJ...677.1216H,2010PhRvD..81l3016H}
\begin{equation}
    H_0'' + {\cal A}_1H_0' + {\cal A}_0 H_0=0\label{H_0_eq},
\end{equation}
where
\begin{eqnarray}
    {\cal A}_0 &\equiv& \nu'' -\frac{6e^{\lambda}}{r^2}-\frac{(\nu')^2}{2} + \frac{3\lambda' + 7\nu'}{2r}\nonumber \\ 
    &-& \frac{\nu'\lambda'}{2} + \frac{\rho'}{P'}\frac{\nu'+\lambda'}{2r}
\end{eqnarray}
and
\begin{equation}
    {\cal A}_1\equiv \frac{2}{r} +\frac{\nu'-\lambda'}{2},
\end{equation}
where $\nu$ and $\lambda$ are related to the metric functions as $g_{tt}\equiv - e^{ \nu(r)}$ and $g_{rr}\equiv e^{ \lambda(r)}$, respectively.

Tidal deformations/deformabilities themselves (dimensionless) are defined as $\Lambda \equiv 2/3(M/R)^{-5}k_2$, where the Love number $k_2$, in terms $y\equiv RH_0'(R)/H_0(R)$, is \citep{2008ApJ...677.1216H,2009PhRvD..80h4035D,2010PhRvD..81l3016H,2011PhRvD..84j3006P}
\begin{eqnarray}
\begin{aligned}
    k_2&=8C^5(1-2C)^2[2+2C(y-1)-y] \\ 
    &{\slash}\big(5\{2C[6-3y+3C(5y-8)]\big.\\ 
    &+4C^3[13-11y+C(3y-2)+2C^2(1+y)] \\
    &+\big.3(1-2C)^2[2-y + 2C(y-1)]\ln(1-2C)\}\big),
    \label{love-number}
\end{aligned}
\end{eqnarray}
with $C\equiv M/R$ is the compactness of the background star. Therefore, it is clear that one needs to find the interior solution to $H_0$ and evaluate it on the surface of the star to obtain $\Lambda$.  

The boundary (interface) conditions that we use here are the continuity of $H_0$ and $H_0'$ at the borders of the mixed state (with the quark and hadronic phases). This is the case since there is no density jumps when the mixed state is taken into account, due to Eqs. \eqref{p_condition} and \eqref{n_condition} and the thermodynamic relation $\mu=(P+\rho)/n$. At the center and on the surface of the star, a regular solution and the absence of energy jumps are considered, respectively (for further details, see \citep{2020arXiv200310781P,2020PhRvD.101j3025G} and references therein). For the tidal deformation calculations in stars with sharp phase transitions, a nontrivial boundary condition at the hadron-quark interface for $H'_0$ should be taken due to the discontinuity of the energy density there (see, e.g., Eq. (41) of \citep{2020arXiv200310781P}).

We stress that Eq. \eqref{H_0_eq} is valid only for perfect fluids in the adiabatic limit. In terms of a binary coalescence, it would be related to the inspiral phase. If parts of the star are elastic, Eq. \eqref{H_0_eq} must be replaced by a set of coupled equations that take into account their shear moduli (see, e.g., \citep{2020arXiv200310781P,2020PhRvD.101j3025G}). Such equations are much more involved and lead to the intuitive result that tidal deformations of elastic stars are smaller than their perfect-fluid counterparts. In most cases, however, the differences are negligible and in general stars with elastic phases have smaller tidal deformations than their perfect-fluid counterparts \citep{2020arXiv200310781P,2020PhRvD.101j3025G}. 
Therefore, even though the mixed state should be elastic, we assume in this work it can be approximated by a perfect-fluid and any allowed EOS in the perfect-fluid case will also be allowed in the more realistic case with elasticity.

\begin{figure*}[t]  
    \includegraphics[width=\textwidth]{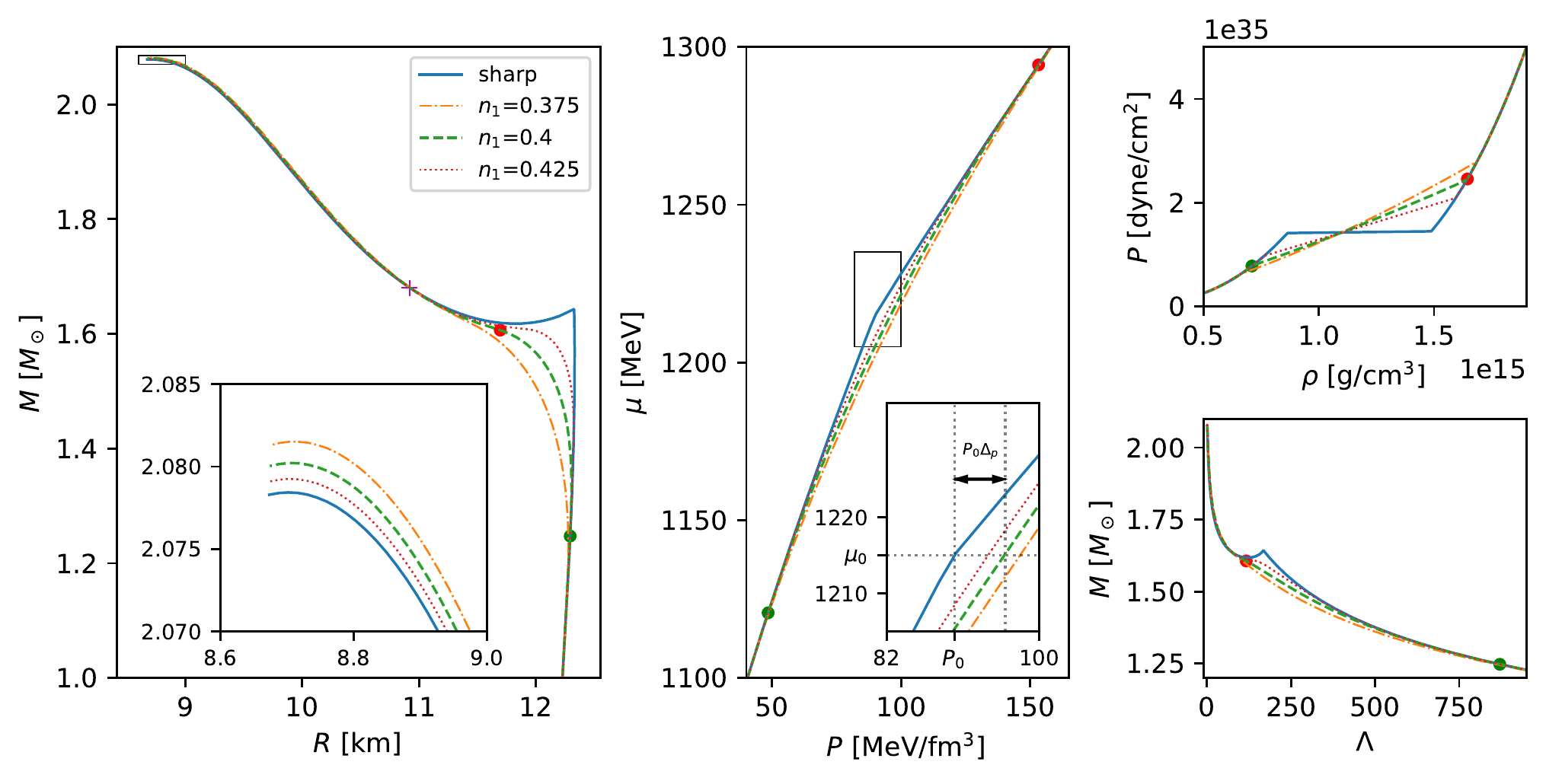}
    \caption{Examples of polytropic EOSs and resulting sequences of NS parameters (solutions of TOV equations): sharp phase transition (solid blue curves), and three mixed state realizations (dash-dotted orange curves with $n_1$=0.375 fm$^{-3}$, dashed green curves with $n_1$=0.4 fm$^{-3}$ and dotted red curves with $n_1$=0.425 fm$^{-3}$). The sharp phase transition EOS parameters are $\gamma_1=3.5$, $\gamma_2=6$, density jump (in terms of the baryon density $n$) between $n_0=0.475$ fm$^{-3}$ and $n_{02}=0.76$ fm$^{-3}$. The leftmost panel contains the mass-radius $M(R)$ sequences (the inset plot presents a closeup of the region around the maximum mass), the middle panel is the chemical potential-pressure $\mu(P)$ relation, the upper right panel is the pressure-density $P(\rho)$ relation, whereas the lower right one is the mass-tidal deformability $M(\Lambda)$ relation. Green and red dots mark the beginning and the end of the mixed-state region in the case of the $n_1$=0.4 fm$^{-3}$ EOS; correspondingly, stellar configurations in the other panels have central parameters equal to the beginning (green dot) and the end (red dot) of the mixed state. The inset in the $\mu(P)$ plot shows the definition of $\Delta_p$ - marked by an arrow - on the example of the $n_1$=0.4 fm$^{-3}$ EOS, marked by the green dashed line. $P(\mu_0)$ is denoted by $P_0$. Note that the for the $M(R)$ sequences the mixed state curves are below the sharp one in the vicinity of the phase transition point, but $M_{\rm max}$ is larger for the mixed state EOSs. For the $n_1$=0.4 fm$^{-3}$ EOS, the mixed state and the sharp transition EOSs have the same mass and radius parameters at $M\approx 1.68\ M_\odot$ and $\approx10.92$ km, marked by a magenta cross.} 
     \label{fig:mixed4}
\end{figure*}

\section{Results}
\label{sect:results} 

Exemplary sequences of sharp- and mixed-state EOSs, based on the polytropic approach of Sect.~\ref{sect:polytropes}, are presented in Fig.~\ref{fig:mixed4}. The parameters of the EOSs are given in its caption. In the following, we will first discuss the origin of the similarity of the $M_{\rm max}$ and $R_{M_{\rm max}}$ for both types of state transitions, and then estimate the sizes of the differences using the EOS approximations from Sect.~\ref{sect:polytropes} and \ref{sect:blaschke}.

\subsection{Equality of $M_{\rm max}$ and other global stellar parameters}
\label{sect:equality_of_mmax} 

The radial dependence of the baryon chemical potential (Gibbs energy per baryon) is obtained from \cite{2017A&A...599A.119Z}:
\begin{equation}
    \frac{d\ln \mu}{dr}=\frac{m}{r^2}\frac{(1+4\pi r^3 P/m)}{1-{2m}/{r}},
    \label{eq:dmudr}
\end{equation}
while the quantity $m$ can be can be calculated using:
\begin{equation}
     \frac{dP}{dm}=\frac{m}{4\pi r^4}\frac{(1+4\pi r^3 P/m)(1+P/\rho)}{1-{2m}/{r}}.
     \label{eq:dpdr}
\end{equation}
For the central pressure larger than the pressure $P_3$, at which a mixed state is fully present in the interior of the star, its region is represented by a shell of thickness $\Delta r_{\rm mixed}=r(P_1)-r(P_3)$ and a mass $\Delta m_{\rm mixed}=m(P_1)-m(P_3)$. Global parameters of this mixed shell, as well as a shell containing the sharp phase transition between ($\mu_1$, $P_1$) and ($\mu_3$, $P_3$) are calculated from Eqs. \eqref{eq:dmudr} and \eqref{eq:dpdr}, and weakly depend on the kind of the EOS (i.e., sharp or  mixed) in this region. Equation~(13) from \cite{2017A&A...599A.119Z} can be used to estimate the thickness of the mixed state region, however it should be stressed that the parameters neglected for the crust ($r^3 P/m$, the change of mass in the considered region) are more important in our case ($P/\rho \simeq 0.1$). As a result, the mass-radius relations for the first order phase transition and the mixed state are almost identical it the regions above ($P_3, \mu_3$). This is exemplified in the next section with the use of many polytropic EOSs.

\subsection{Results for polytropic EOSs}
\label{sect:polytropes_results}

In order to decide which cases might be observationally relevant, we make use of already existing and future mass, radius and tidal deformation measurement accuracies. NICER measurements already allow the constraint of masses and radii of NSs with relative uncertainties around $5\%$ (for combined observations), and around $10\%$ for single observations (see, e.g., \citep{2021ApJ...918L..28M} and references therein) at 1$\sigma$ level. Future missions, such as eXTP or Athena are expected to measure the above quantities with even smaller fractional uncertainties, around a few percent (say, $1\%-2\%$ in the most optimistic cases; at the $90\%$ credible interval, uncertainties would increase accordingly). When it comes to tidal deformations, current relative uncertainties are still large ($\sim 50\%-100\%$), but future measurements (e.g., with third generation GW detectors) could deliver uncertainties as small as $2\%$ at the 90$\%$ credible level in the most optimistic cases \citep{2021arXiv210812368C}. For 2.5-generation detectors and less optimistic cases, relative uncertainties of $5\%-10\%$ are expected \citep{2021arXiv210812368C}. When translated to radius constraints, they could also be around a few percent ($1\%-2\%$) for the most optimistic cases at the same credible level as above \citep{2021arXiv210812368C}. 

\begin{figure} 
   \includegraphics[width=\columnwidth]{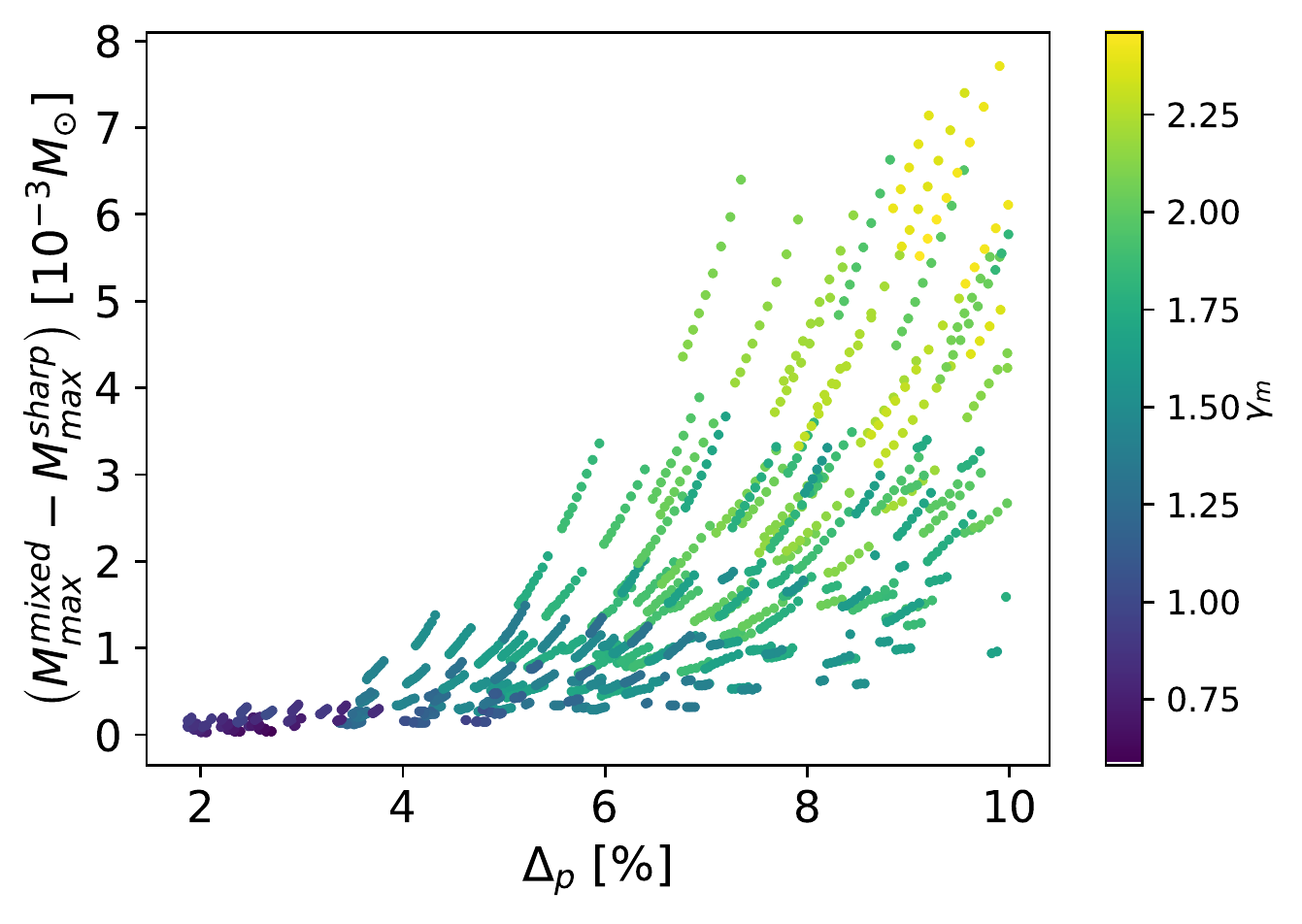}   
   \caption{Mass differences of hybrid stars with mixed states and sharp phase transitions for the maximum masses for several polytropic equations of state as a function of $\Delta_p$.} \label{fig:DeltaM_Deltap_pol}
   \end{figure}

\begin{figure} 
   \includegraphics[width=\columnwidth]{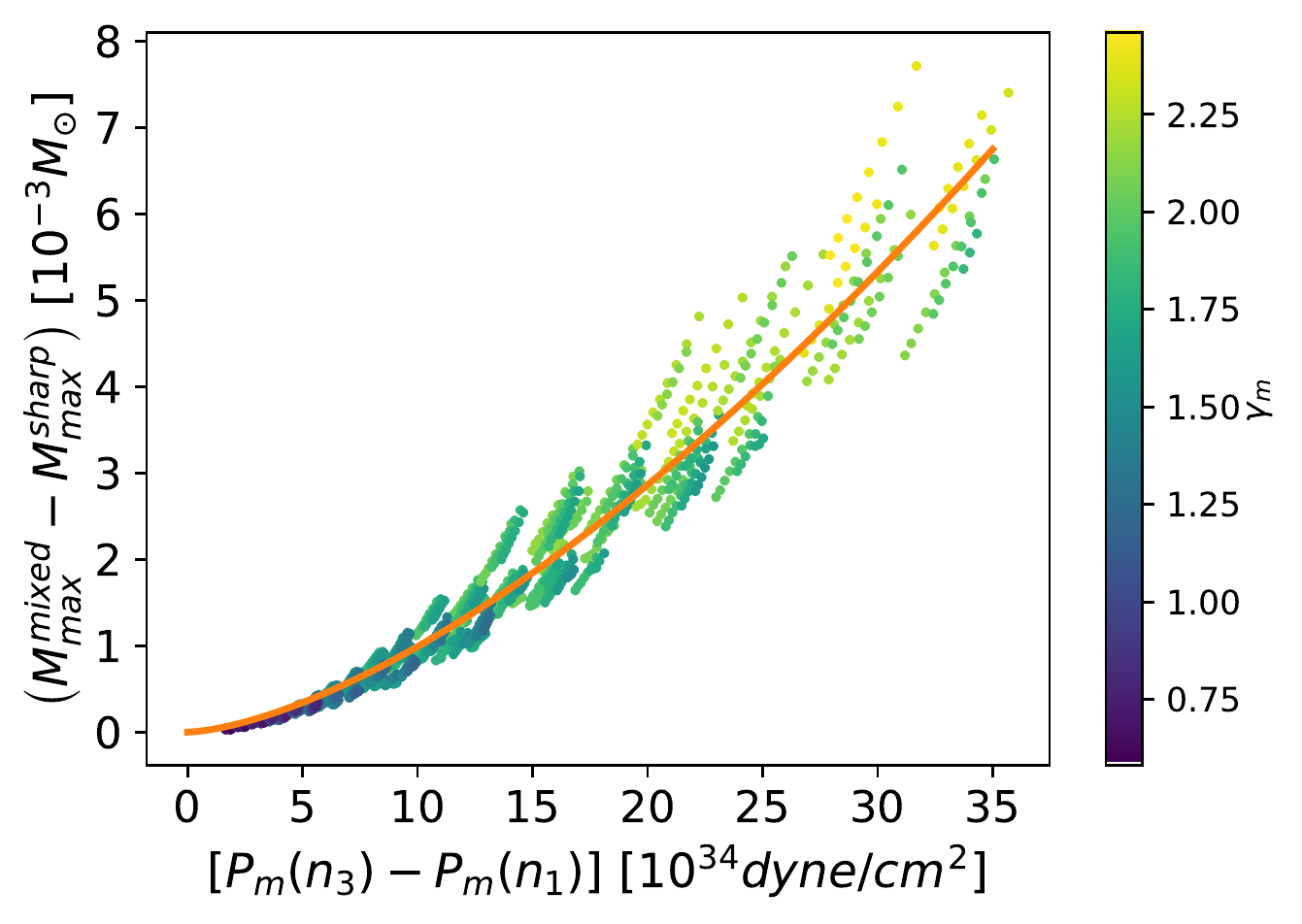}   
   \caption{Same as Fig. \ref{fig:DeltaM_Deltap_pol} but now taking into account the pressure difference between the bottom and the top of the mixed state. The fit in the plot has the form $y=a_px^p$, with $a_p=2.911\times 10^{-2}$ and $p=1.5317$.  } \label{fig:DeltaM_pres_pol}
   \end{figure}

\begin{figure} 
   \includegraphics[width=\columnwidth]{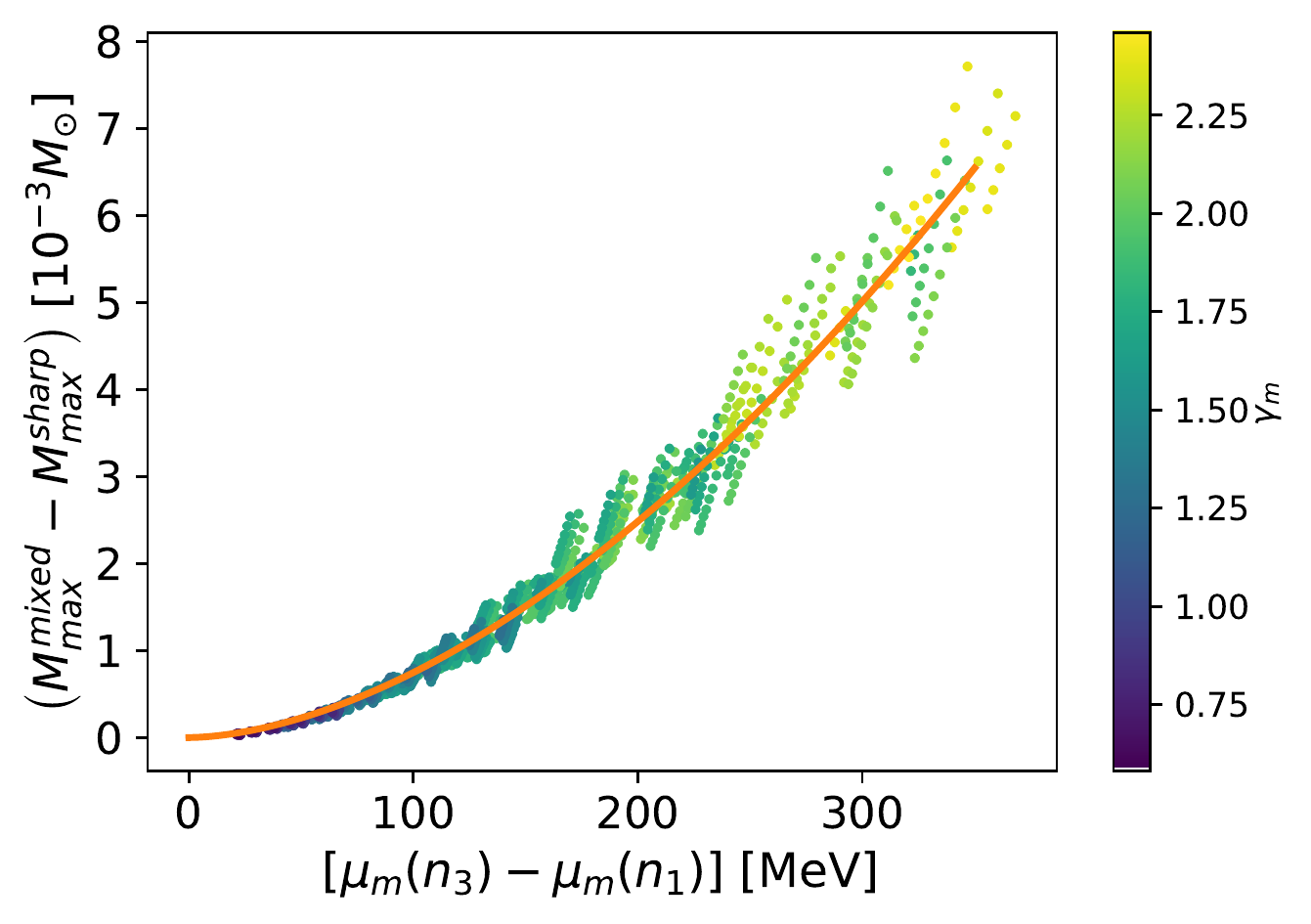}   
   \caption{Maximum mass dependence on the chemical potentials at the borders of the mixed state. The power-law fit of Fig. \ref{fig:DeltaM_Deltap_pol} gives $a_p=2.573\times 10^{-4}$ and $p=1.7316$. } \label{fig:DeltaM_frac_pres_pol}
   \end{figure}

\begin{figure} 
   \includegraphics[width=\columnwidth]{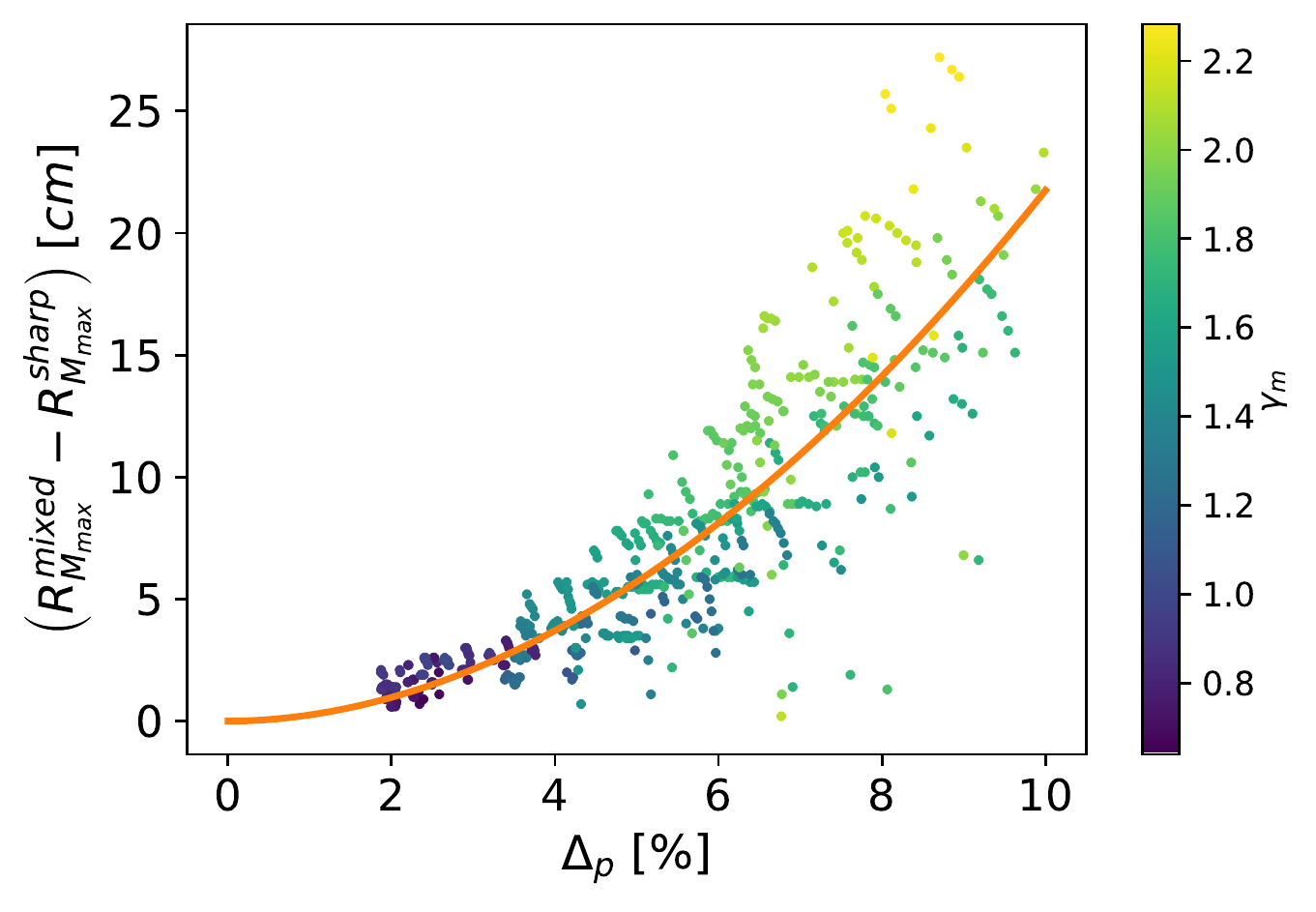}   
   \caption{Radius differences associated with the maximum masses for stars with mixed states and sharp phase transitions for several polytropic equations of state as a function of $\Delta_p$. The parameters for the power-law fit ($y=a_px^p$) are $a_p=0.2557$ and $p=1.930$.
   } \label{fig:DeltaR_delta_p_pol}
   \end{figure}

The effectiveness of the approximation from Sect.~\ref{sect:equality_of_mmax} is demonstrated by  comparing a large set of sharp phase transition EOSs with their corresponding mixed-state EOSs. The prescription is based on the polytropic approximation of Sect.~\ref{sect:polytropes}, where the values of polytropic indices and baryon densities, 
denoting the beginnings and ends of the phases, cover the following ranges: $\gamma_1\in (2.75, 3.75)$, $\gamma_2\in (4.5, 6.5)$, $n_{0} \in (0.4, 0.5)\,{\rm fm^{-3}}$, $n_{02}/n_{0} \in (1.45, 1.65)$ (we only consider here strong phase transitions i.e. large quark-hadron density jumps in order to maximize observable differences), $n_1 \in (0.325, 0.4)\,{\rm fm^{-3}}$; $\gamma_m$ and $n_3$ were solutions to the thermodynamic conditions and roughly varied in the intervals $(0.5,2.5)$ and $(0.5, 1.0)\,{\rm fm^{-3}}$, respectively. These parameter intervals lead to observationally reasonable NS parameters, and also reflect expectations regarding the densities phase transitions might take place in stars (see \citep{2019A&A...622A.174S} for further details). For final comparisons, we select only  those microscopic models that lead to $M_{\rm max} > 2\,M_\odot$, $\mu_m(n_3)/\mu_m(n_{02})<1.15$, $\mu_m(n_1)/\mu(n_0)>0.85$ (see  \cite{2019PhRvC.100b5802M} for the reasonableness of these limits). In addition, we take into account the tidal deformability and mass constraints coming from GW170817 \citep{2019PhRvX...9a1001A}.\footnote{We do not take into account all current observations because that would be beyond the goal of the present paper. They already have been done elsewhere and one of the outcomes is that strong phase transitions are fully allowed \cite{2021ApJ...913...27L}. Even if the radius outcome of PSR J0740+6620  ($M=2.08\pm0.07$~$M_{\odot}$) is taken into account \citep{2021ApJ...918L..28M}, the possibility of strong phase transitions would not be affected because the stiff EOSs used in \cite{2021ApJ...913...27L} are already compatible with such radius. 
We leave NICER constraints on NS radii (e.g., \citep{2019ApJ...887L..24M,2021ApJ...918L..28M}) within our analysis for future work. In it, also weak phase transitions will be taken into account, which could be done simply by changing the parameters of our polytropic model. For both strong and weak phase transitions, though, {\it relative changes} in global NS parameters should not be greatly affected by NS constraints because they mostly depend on mixed state properties, which change little masses and radii of NSs (see Sec. \ref{sect:polytropes}). Indeed, relative changes -- due to our assumption of the same low- and high-density EOSs -- almost disregard these contributions, and they are exactly the ones that would mostly affect radii and masses of stars.} We keep all $\gamma_m$ fulfilling the above conditions in order to better explore the region of parameters of the mixed state and also to check consistency, given that low mixed phase adiabatic indices would be a rough proxy for sharp phase transition EOSs.
Figures \ref{fig:DeltaM_Deltap_pol}, \ref{fig:DeltaM_pres_pol} and \ref{fig:DeltaM_frac_pres_pol} show the differences (sharp phase transitions and mixed states) for the maximum mass, and Fig. \ref{fig:DeltaR_delta_p_pol} shows the radius differences at the maximum mass, as a function of several parameters: $\Delta_p$, $P_m(n_3) - P_m(n_1)$ and $\mu_m(n_3)-\mu_m(n_1)$. Each point in Figs. \ref{fig:DeltaM_Deltap_pol}, \ref{fig:DeltaM_pres_pol}, \ref{fig:DeltaM_frac_pres_pol} and \ref{fig:DeltaR_delta_p_pol} is the result of the comparison of two EOSs (sharp phase transition and mixed state) that only differ in the pressure interval $P_1<P<P_3$. For illustration and clarity, we have color-marked only the $\gamma_m$s of the mixed state EOSs. One can clearly see that the fractional differences are very small: around $10^{-1}\%$ for the maximum mass and $10^{-3}\%$ for the associated radius. The larger scatter in Fig. \ref{fig:DeltaR_delta_p_pol} is simply due to the general flattening of the mass-radius relation around the maximum mass. In general, smaller values of $\gamma_m$ lead to smaller differences, as consistency would demand. 

   %%%%%%%%%%%%%%%%%%%%%%%%%%%
   
   %%%%%%%%%%%%%%%%%%%%%%%%%%%

   %%%%%%%%%%%%%%%%%%%%%%%%%%%
   
      %%%%%%%%%%%%%%%%%%%%%%%%%%%

\begin{figure} 
   \includegraphics[width=\columnwidth]{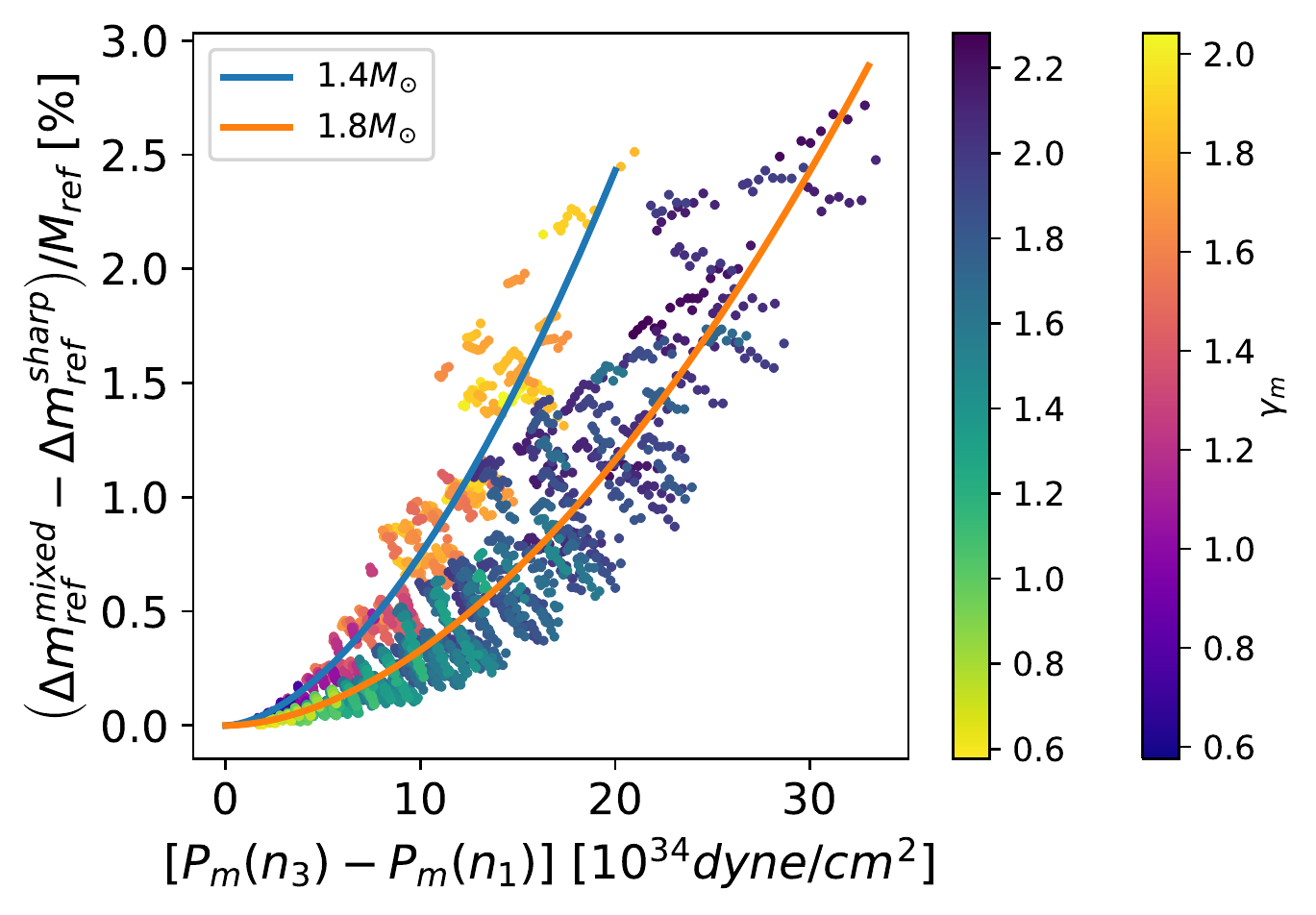}   
   \caption{Masses of
the shells containing mixed states
 (within pressures $P_1$ and $P_3$) subtracted by masses encompassed within the same pressures in stars with sharp phase transitions, normalized by the reference (``ref'') masses (either 1.4 or 1.8 solar masses). For 1.4 $M_\odot$ stars,
 the fit ($y=a_px^p$) is such that $a_p=1.506\times 10^{-2}$ and $p=1.697$.  For 1.8 $M_{\odot}$, it follows that 
$a_p=5.020\times 10^{-3}$ and $p=1.818$.
   } \label{fig:DeltaM_thickness_pol}
   \end{figure}

\begin{figure} 
   \includegraphics[width=\columnwidth]{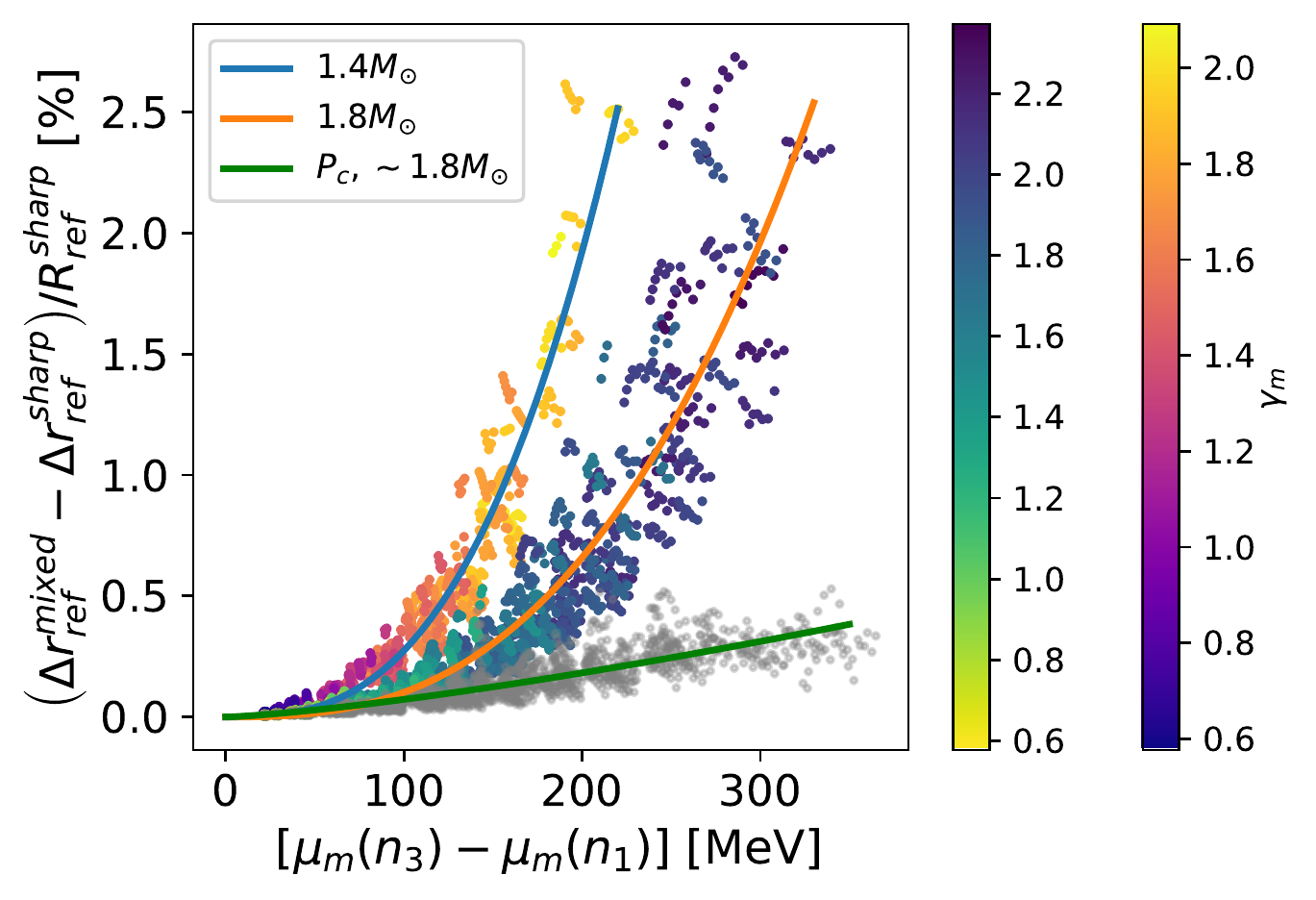}   
   \caption{Thickness differences of the mixed states of stars with respect to sharp phase transitions for given reference masses (and for completeness the same central pressure $P_c$ for a sharp EOS and its associated mixed EOSs), normalized by the radius of stars with sharp phase transitions ($R^{\rm sharp}_{\rm ref}$). For 1.4 $M_\odot$ stars, the fit $y=a_px^p$ implies that $a_p=6.509\times 10^{-7}$ and $p=2.812$. For 1.8 $M_{\odot}$, it follows that
   $a_p=4.185\times 10^{-7}$ and $p=2.693$. When the central pressure of a star with a sharp phase transition and another one with a mixed state is the same (their masses are not the same; we choose $P_c$s in stars with mixed-state EOSs so their masses are 1.8 $M_{\odot}$), we have that $a_p=1.622\times 10^{-4}$ and $p=1.326$.
   } \label{fig:DeltaR_thickness_pol}
   \end{figure}

\begin{figure} 
   \includegraphics[width=\columnwidth]{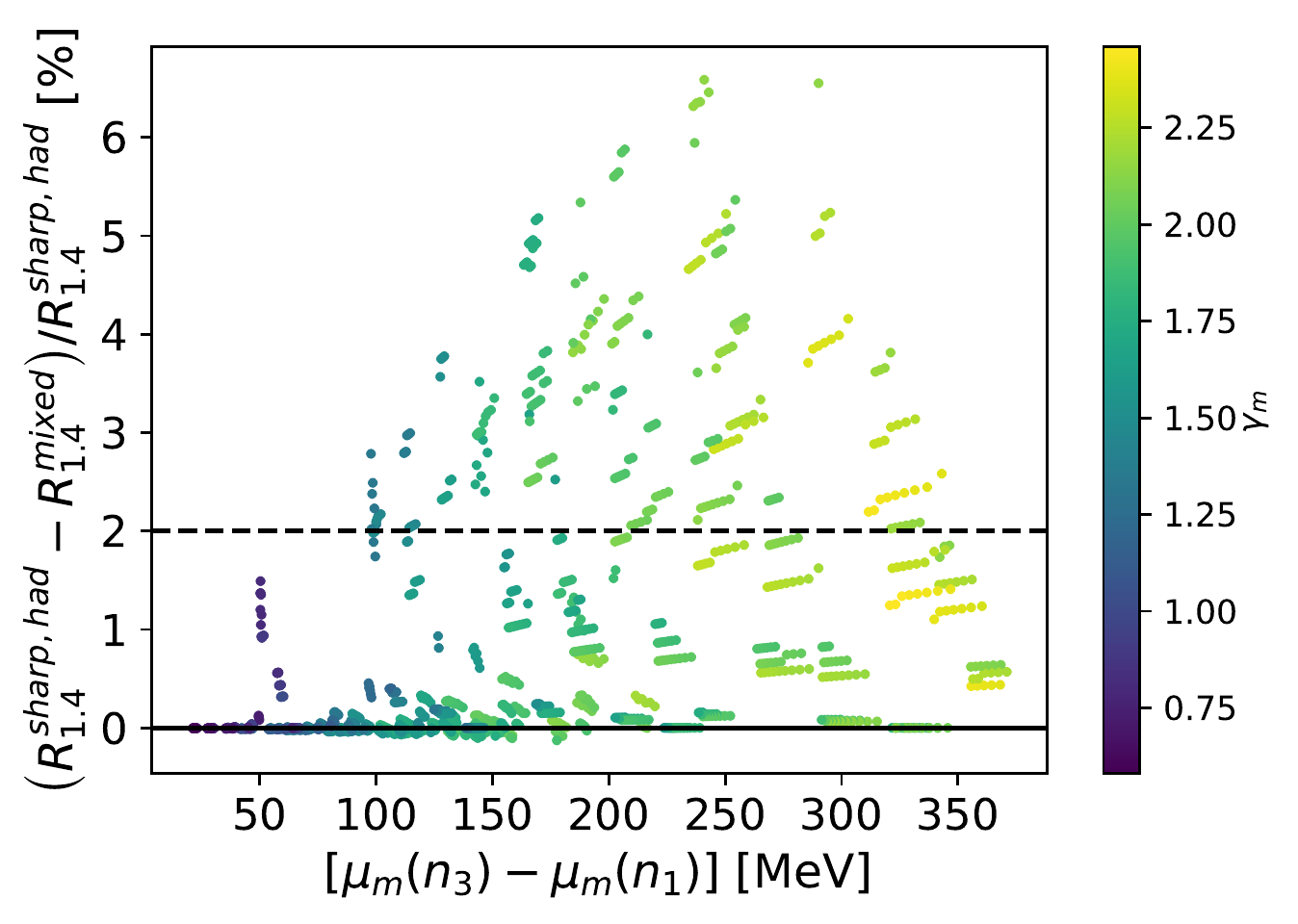}   
   \caption{ Relative radius differences for 1.4 $M_{\odot}$ stars with mixed states, sharp phase transitions and even purely hadronic (``had'')/one-phase. Here, $2\%$ is a representative level of accuracy of near-future electromagnetic/GW missions. The largest differences are connected with the comparison of one-phase stars with those with mixed states. The smallest differences ($\lesssim (1-2)\%$), though, come from stars with mixed states and those with sharp phase transitions and quark cores.
   } \label{fig:radius_pol_1_dot_4}
   \end{figure}

\begin{figure}
   \includegraphics[width=\columnwidth]{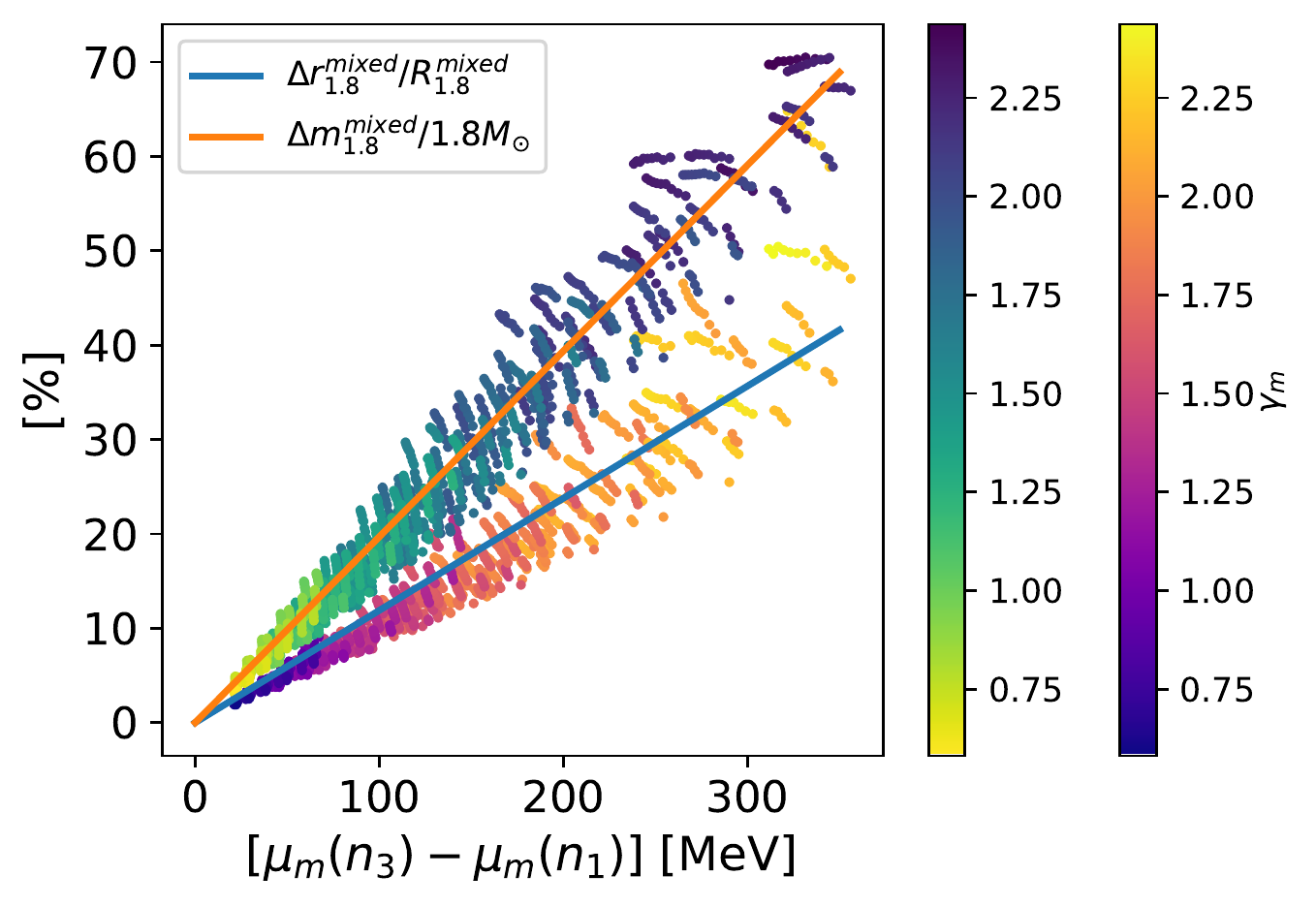}   
   \caption{Relative thicknesses and masses of mixed states in hybrid stars with 1.8 $M_{\odot}$. Straight lines ($y=a_1 x$) fit well the data; for the relative thickness size of the mixed state, $a_1=0.1189$; for its relative mass, $a_1=0.1969$. The size and mass of the mixed state is non-negligible when compared to the corresponding aspects of the star in general, and it increases with the increase of $\gamma_m$.
   } \label{fig:mass_radius_mixed_pol}
   \end{figure}

Regarding $\Delta r_{\rm mixed}$ and $\Delta m_{\rm mixed}$, Figs. \ref{fig:DeltaM_thickness_pol} and \ref{fig:DeltaR_thickness_pol} show how aspects of a mixed state compare with aspects of a sharp phase transition for a region between $P_1$ and $P_3$ in the case of hybrid stars with the same mass (usually different central pressures), taken representatively here as 1.4  and 1.8 $M_{\odot}$. Naturally, in order to do so, we have only taken stars whose $P_3$ are smaller than their central pressures for given reference masses. One can see that for almost all cases, fractional changes of the mass and the thickness for sharp phase transitions and mixed states are at most of a few percent, and the difference decreases, for a given $P_m(n_3)-P_m(n_1)$, when the mass of the star increases. (That would qualitatively explain why the differences are so small for the maximum masses and associated radii of stars.) The color maps also make it clear that differences between mixed-state and sharp EOS aspects increase with $\gamma_m$. This is reasonable given that a mixed-state EOS becomes harder for larger $\gamma_m$. When it comes to the fractional radius differences for a given mass, Fig. \ref{fig:radius_pol_1_dot_4} for 1.4 $M_{\odot}$ (stable) stars suggests that in some cases the differences could be observed by near-term detectors. The largest differences concern purely hadronic (one-phase) stars (whose phase-transition masses are larger than 1.4 $M_{\odot}$) being compared with stars presenting mixed states. If, instead, one restricts the comparison to 1.4 $M_{\odot}$ stars all having quark cores, the radius changes decrease significantly, and are up to around 1$\%$ - 2$\%$. Similar or even smaller upper-limit relative changes to the radii also come for NS masses the range of $\sim (1-2)M_{\odot}$. Thus, it suggests that the minimum radius precision detectors should have in order to start differentiating mixed states from sharp phase transitions in general is indeed around $1\%-2\%$. For the mass precision, any small region of the $M(R)$ diagram could be interpolated as $M-M_{\rm ref}={\rm const}\times(R-R_{\rm ref})$ (``ref'' stands for a reference value), meaning that $\Delta M/M =|(1-M_{\rm ref}/M)/(1-R_{\rm ref}/R)|\Delta R/R$. For instance, in the example of Fig. \ref{fig:mixed4} for the hybrid branch, we have that $|(1-M_{\rm ref}/M)/(1-R_{\rm ref}/R)|\sim (0.8-1.5)$. Thus, relative mass accuracy should closely trail the radius accuracy.
All the above also shows that mixed states and sharp phase transitions between $P_1$ and $P_3$ almost share the same macroscopic properties, despite being very different physically and encompassing non-negligible portions and masses to stars in general. Indeed, the thicknesses occupied by mixed states and their masses could be even larger than half of a star radius and a third of its total mass, respectively,
as shown in Fig. \ref{fig:mass_radius_mixed_pol} for 1.8 $M_{\odot}$ stars.

%%%%%%%%%%%%%%%%%%%%%%%%%%%

%%%%%%%%%%%%%%%%%%%%%%%%%%%

%%%%%%%%%%%%%%%%%%%%%%%%%%%

%%%%%%%%%%%%%%%%%%%%%%%%%%%

The similarity between mixed state and sharp phase transition regions in stars can be theoretically explained on basis of Eqs. \eqref{eq:dmudr} and \eqref{eq:dpdr}, and it could be more clearly seen for stars with the same central pressure (larger than $P_3$). In this case, stellar configurations with mixed states and sharp-phase transitions have the same boundary conditions at $P_3$ ($r(P_3)$, $m(P_3)$), and the thickness and the mass of the region between $P_1$ and $P_3$ is approximately given by $\Delta\mu_m=\mu_3-\mu_1$ and $\Delta P_m=P_3-P_1$, respectively. As a result, we get that the difference of the thickness of regions between $P_1$ and $P_3$ for a sharp-phase-transition and a mixed-state star with a same central pressure is much smaller than when they have the same mass. This could be seen in particular in Fig. \ref{fig:DeltaR_thickness_pol} for stars with masses around and exactly $1.8M_{\odot}$ (green and orange curves, respectively), where their maximum fractional changes differ by a factor of $\sim 5$. Notwithstanding, we have found that the maximum relative changes of radii of stars with sharp phase transitions and mixed states with the same central pressure are roughly similar to those with the same mass.

\begin{figure} 
   \includegraphics[width=\columnwidth]{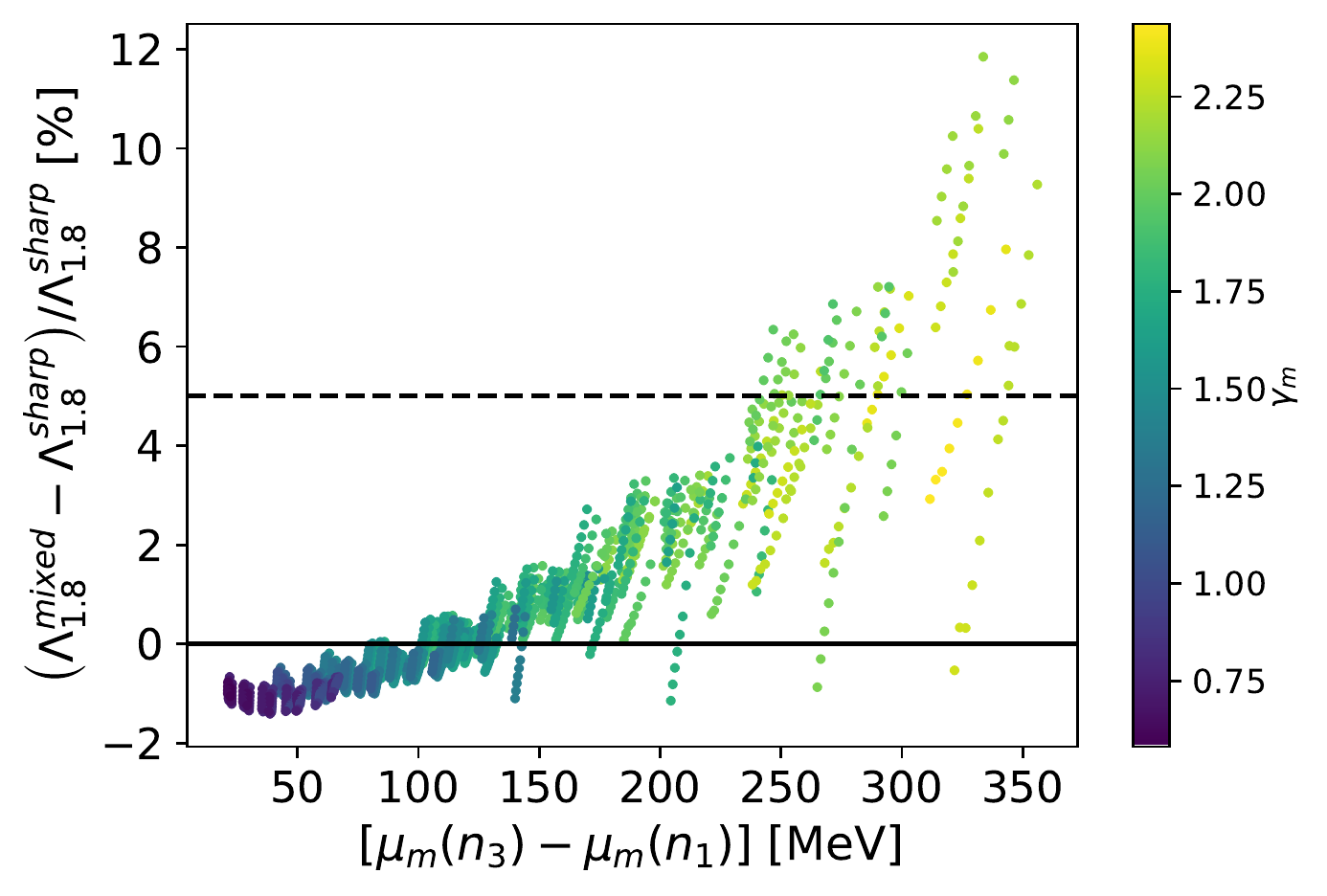}   
   \caption{Relative tidal deformabilities of 1.8$M_{\odot}$ hybrid stars (with mixed states and sharp phase transitions and quark cores) as a function of the chemical potential difference at the beginning and the end of the mixed state. The threshold of around $5\%$ precision is reached for chemical potential differences larger than approximately $250$ MeV and $\gamma_m\gtrsim 1.5$.
   } \label{fig:tidal_pol}
   \end{figure}

\begin{figure} 
   \includegraphics[width=\columnwidth]{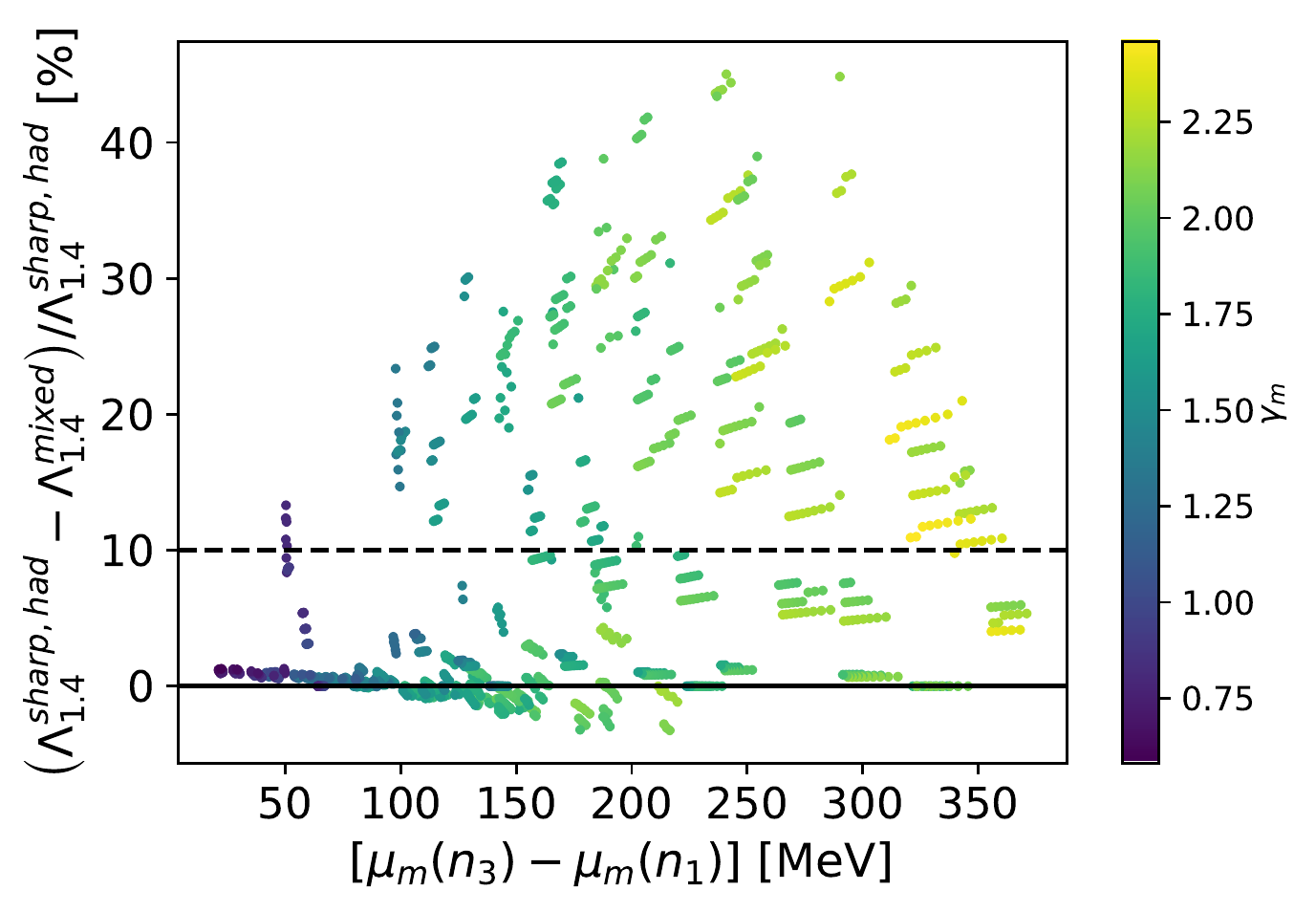}   
   \caption{Relative tidal deformabilities of 1.4$M_{\odot}$ stars relaxing the constraint that they only have quark cores and central pressures larger than the ones fully encompassing the mixed state: here they can also be purely hadronic (``had'') or have mixed states at their centers. The tidal deformability normalization has been chosen to be those of 1.4 $M_{\odot}$ stars either presenting sharp phase transitions or being purely hadronic, which depends on the phase transition masses due to their EOSs. The largest differences are obtained when one compares an one-phase star with another one having a mixed state (for a same mass), for large values of $\gamma_m$. Fractional differences even exceeding $10\%$ could happen for several EOSs and $\gamma_m$s. When stars only with sharp phase transitions and mixed states are compared, maximum tidal deformations are of a few percent ($\lesssim 2\% - 3\%$).
   } \label{fig:tidal_pol_broader}
   \end{figure}

When tidal deformations are concerned, fractional changes could be much larger and could exceed the rough threshold of detectability for 3G GW detectors (uncertainties as small as $2\%$) for certain cases, as clear from Fig. \ref{fig:tidal_pol} for 1.8 $M_\odot$ hybrid stars (naively speaking, they are more likely to have quark cores due to larger central pressures than $1.4 M_{\odot}$ stars and also are more likely to be detected than $2M_{\odot}$ stars in terms of tidal deformations). Indeed, the above figure suggests that optimist minimum tidal deformation precisions for starting differentiating sharp-phase transitions from mixed states would roughly be $5\% - 10\%$. This is also roughly the case for hybrid stars with masses $\sim (1.5 - 1.8) M_{\odot}$. Slightly higher upper limits arise for masses larger than 1.8 $M_{\odot}$ due to the possibility of larger chemical potential differences at the base and top of the mixed state. For masses smaller than $\sim 1.5 M_{\odot}$, due to the smaller chemical potential range for the mixed state, relative tidal deformation thresholds should be smaller than a few percent, as also evidenced by Fig. \ref{fig:tidal_pol_broader} for 1.4 $M_{\odot}$ when one only focuses on hybrid stars. When purely hadronic stars are also compared with those having mixed states for a same mass, relative tidal deformabilities $(1-\Lambda^{\rm mixed}/\Lambda^{\rm had})$ as high as around $40\%-50\%$ could emerge, as clear from Fig. \ref{fig:tidal_pol_broader} for 1.4 $M_{\odot}$. This suggests that observationally identifying a softening of the EOS -- due to a phase transition -- would in principle be simpler and it should happen before we might be able to start differentiating a sharp phase transition from a mixed state, exactly as we have assumed.

From all the above, tidal deformations may be a relevant observable for distinguishing sharp phase transitions from mixed states in NSs. We come back to this issue later on. The above figures also reveal that depending on where in mass the quark phase appears, the sharp phase transitions could lead to either larger or smaller tidal deformations than mixed states. Therefore, for a given mixed-state EOS, there should exist a critical mass (a ``crossing mass'') above which tidal deformations of stars with mixed states are larger than their sharp-phase-transition counterparts, and do not always chase the latter down from below. Figures  \ref{fig:mixed4} and \ref{fig:radius_pol_1_dot_4} also show some aspects of this crossing, which happens at different masses in the $M-R$ and $M-\Lambda$ diagrams, for a particular EOS.  

%%%%%%%%%%%%%%%%%%%%%%%%%%%
%%%%%%%%%%%%%%%%%%%%%%%%%%%%%%%%%%%%%%

%%%%%%%%%%%%%%%%%%%%%%%%%%%%%%%%%%%%%%

In our set of polytropic EOSs, accidentally, we have not reached very small values for $\Delta_p$ (or any other parameter difference between the top and the bottom of the mixed state), due to the particular combination of parameters required for that. However, it is more controllable to do so using the parabolic construction of \citeauthor{2018Univ....4...94A}. As the plots in the next section will show, when the mixed state is very thin ($\Delta_p\rightarrow 0^+$), the observables converge to the sharp-phase transition ones. That is very clear from the EOS point of view (because they are basically the same), which is the basis for any observable.

\subsection{Results for the \citeauthor{2018Univ....4...94A}~\citeyear{2018Univ....4...94A} construction}

The analysis of the large set of polytropic mixed-state EOSs from the previous section has given us many clues on the relevant cases to focus on in terms of observations. Here we particularize the analysis to the (parabolic) mixed state construction of \citeauthor{2018Univ....4...94A} to show that different constructions roughly agree among themselves and
to shed some light on those parametric results. As a result of this agreement, the conclusions about radius and tidal deformation accuracies needed to start differentiating a mixed state from a sharp phase transition as drawn before should be relatively free of mixed-state EOS aspects. Another consequence of this agreement would be that in future analyses one might choose only a model to work with.

We have considered two examplary EOSs for sharp phase transitions with different density jumps $\eta\equiv \rho_-/\rho_+-1$, and use them to construct the mixed state following the parabolic prescription \citeauthor{2018Univ....4...94A} 2018 for different $\Delta_p$s.
Here we have defined $\rho_-(\rho_+)$ as the density at the top(base) of the quark(hadron) phase in the case of the Maxwell construction. The first EOS concerns a sharp strong phase transition\footnote{Strong phase transitions do not meet the Seidov criterion for a stable hybrid star, $\eta<1/2(1+P_{0}/\rho_+)$ \citep{1971SvA....15..347S}, whereas weak ones do. A Maxwell construction not fulfilling the Seidov criterion would not lead to stable stars with infinitesimally small quark cores \citep{2008A&A...479..515Z}, meaning that the hybrid $M(R)$ branch is not continuously connected with the purely hadronic one. The minimum size of the quark core is determined through the condition $\partial M /\partial R = 0$ around the phase transition mass.} and we take the EOS of Fig.~\ref{fig:mixed4} as a reference. Just for completeness, the second EOS is related to a sharp weak phase transition (see, e.g., \citep{2019A&A...622A.174S} and references therein). 
For simplicity and convenience, we choose a simple bag-like model for the quark core with $c_s^2=1$. 
This is done because it is the stiffest EOS for the core and hence would in general maximize the stellar parameters and hence the differences between sharp and mixed-state transition outcomes. It is joined to a polytropic EOS for the inner crust and then, around and below (smaller densities) the nuclear saturation density, the SLy4 EOS. For further details, see \cite{2020arXiv201106361P}.

In our forthcoming analysis we will focus on the minimum relative uncertainties for radius and tidal deformabilities suggested by our polytropic studies and check if they are also benchmarks for different mixed-state constructions. For the reference values, we mostly take the ones close to the appearance of the quark phase in the Maxwell construction since they maximize the departures from a mixed state and a sharp phase transition. 

\begin{figure} 
   \includegraphics[width=\columnwidth]{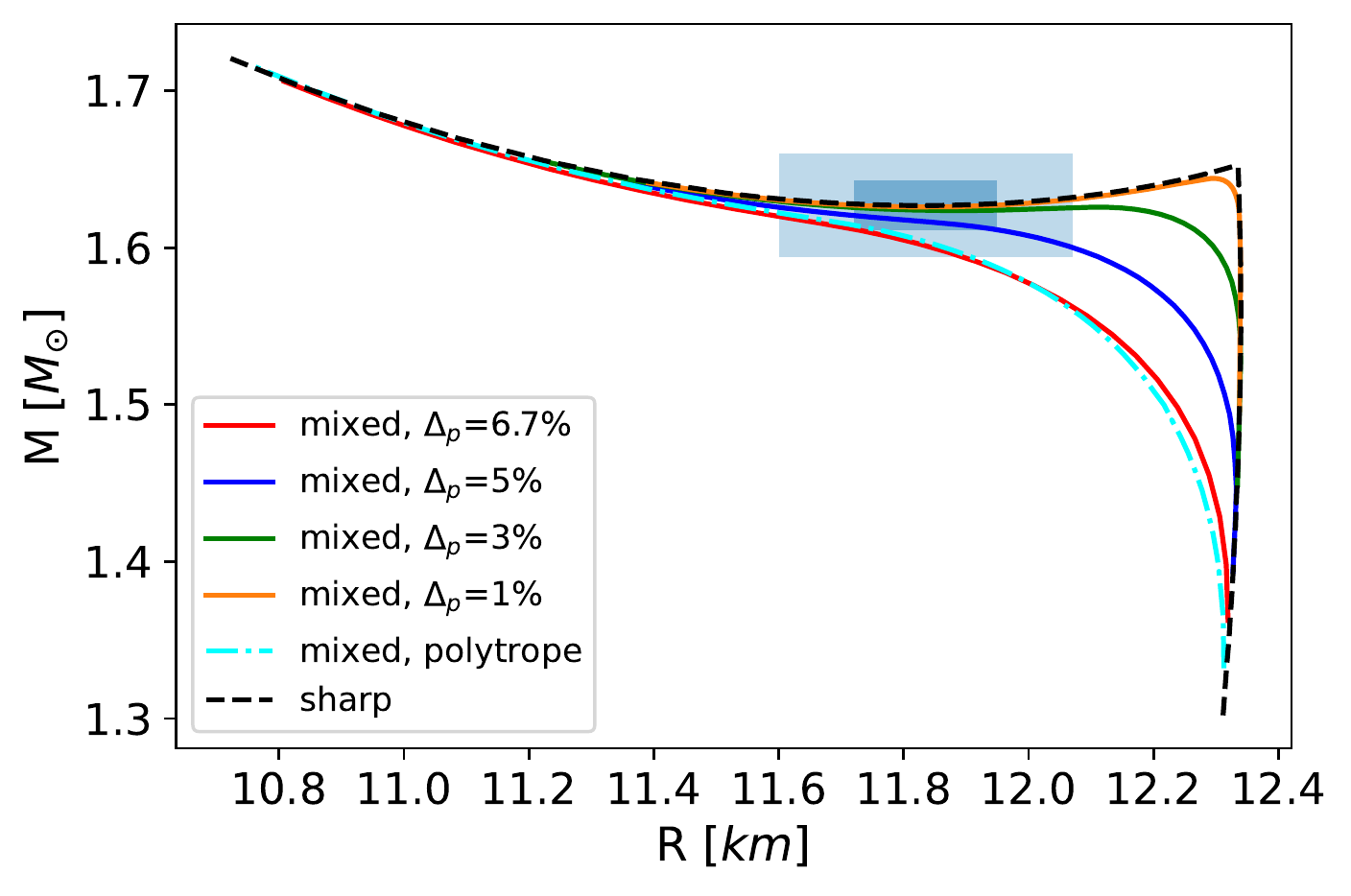}   \caption{Mass-radius relations around the appearance of a quark phase for stars with and without the mixed state for $\eta=0.71$ (same sharp-phase-transition EOS of Fig. \ref{fig:mixed4}). The dot-dashed cyan curve is a zoom-in of the correspondent curve in Fig. \ref{fig:mixed4} for a polytropic construction of the mixed state ($n_1=0.4\, {\rm fm}^{-3}$ EOS) around the appearance of the quark phase. The darker (lighter) box corresponds to masses and radii with (representative) $1\%$ ($2\%$) fractional uncertainties, centered around the critical point for the Maxwell construction ($M={1.63{ M}_{\odot}}$, $R=11.84$ km).}\label{M_R_071}
   \end{figure}

\begin{figure} 
   \includegraphics[width=\columnwidth]{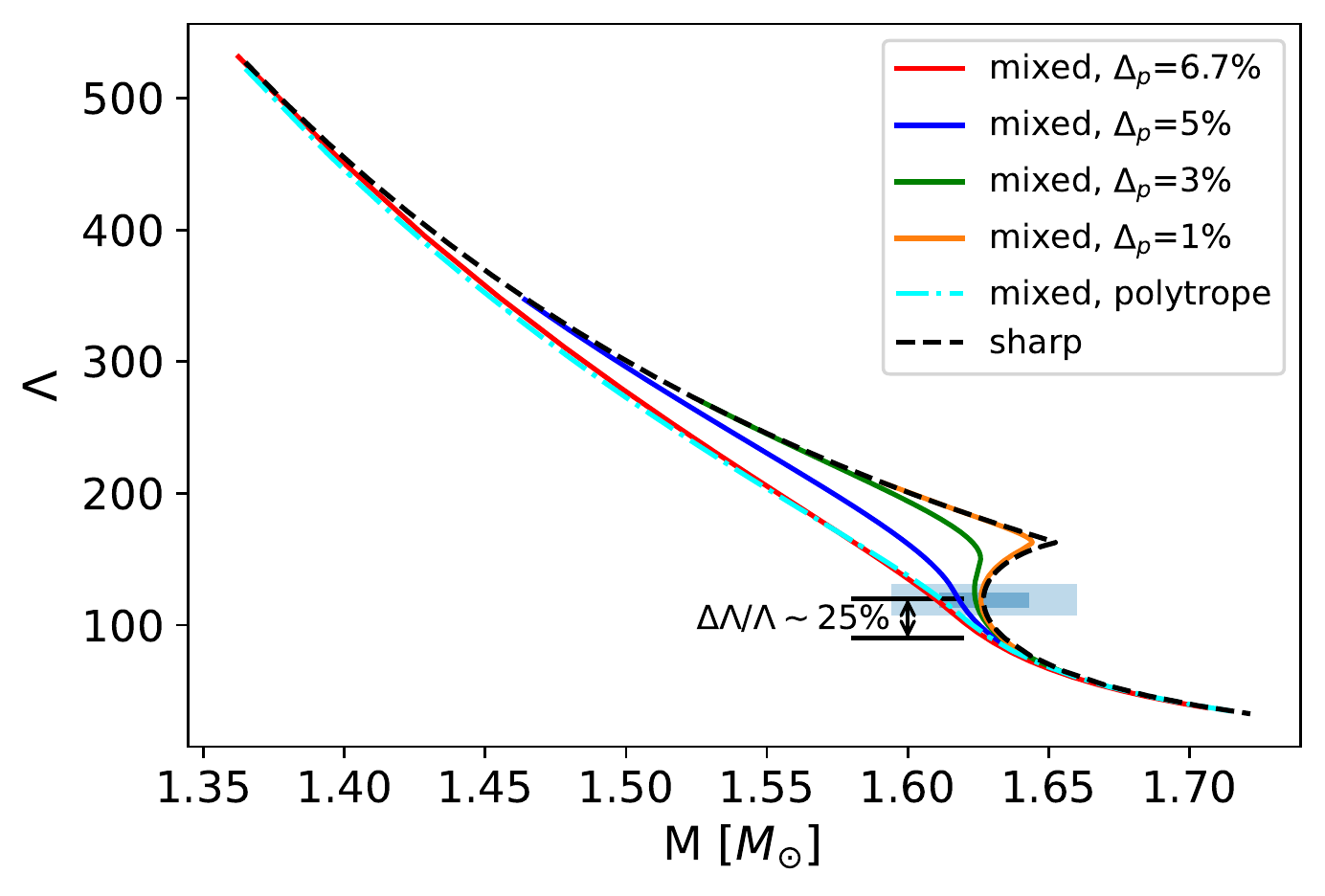}   
   \caption{Mass-tidal deformability relations around the appearance of a quark phase for stars with and without the mixed state for $\eta=0.71$. The dot-dashed cyan curve is also a zoom-in of the correspondent plot in Fig. \ref{fig:mixed4} for the $n_1=0.4\, {\rm fm}^{-3}$ EOS. The darker (lighter) box corresponds to tidal deformability with (representative) $5\%$ ($10\%$) fractional uncertainties. The mass uncertainties are the same of Fig.\ref{M_R_071}. Around the onset of stable quark phases ($M= 1.63 M_{\odot}$ and $\Lambda= 120$), relative tidal deformability differences associated with strong sharp phase transitions and mixed states could be up to around $25\%$. }\label{M_tidal_deformation_0805}
   \end{figure}

\begin{figure} 
   \includegraphics[width=\columnwidth]{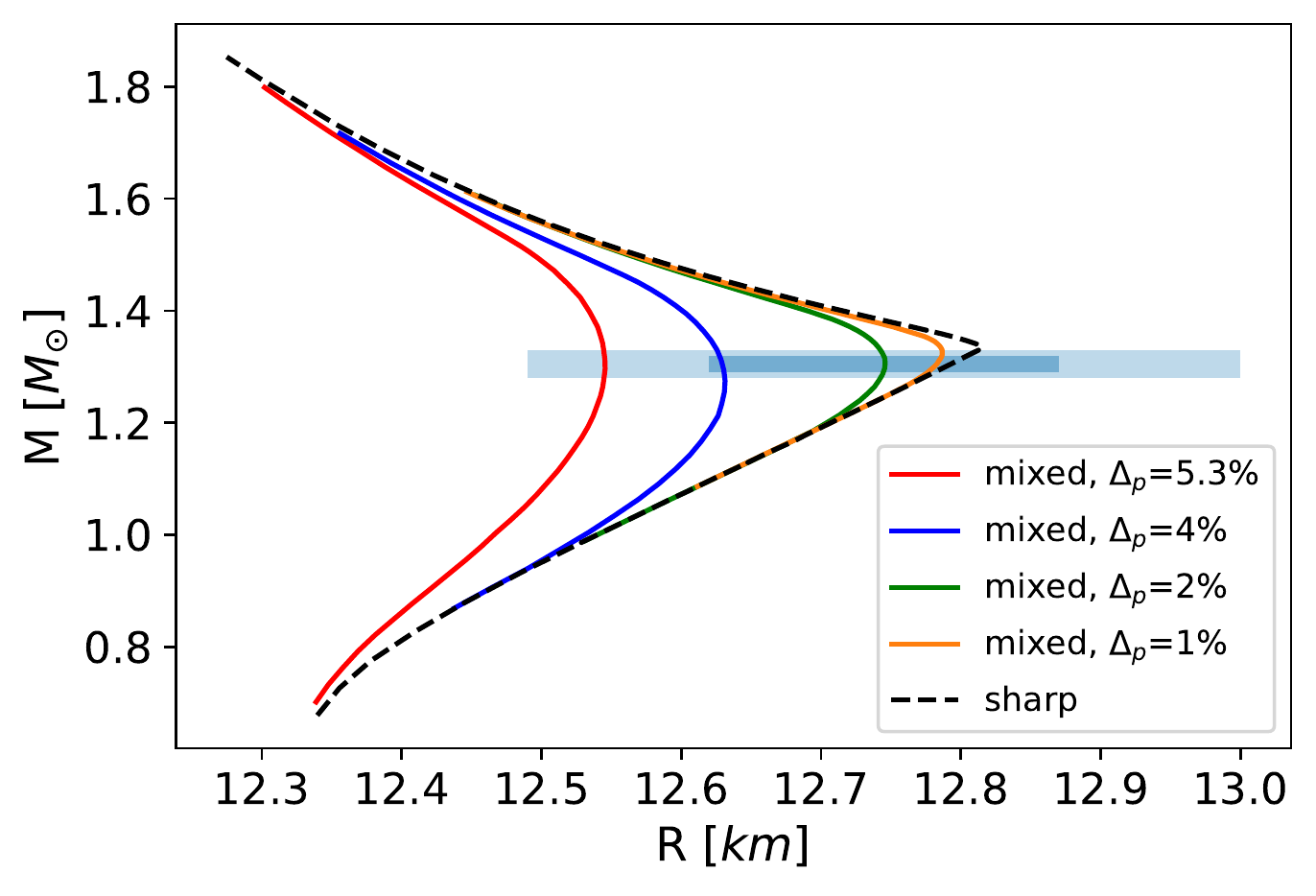}   \caption{Same as Fig. \ref{M_R_071} but for $\eta=0.391$ (weak phase transition). We centered uncertainty boxes at $M=1.31M_{\odot}$, $R=12.75$ km (inflection point of the $\Delta_p=2\%$ curve).}\label{M_R_0391}
   \end{figure}

\begin{figure} 
   \includegraphics[width=\columnwidth]{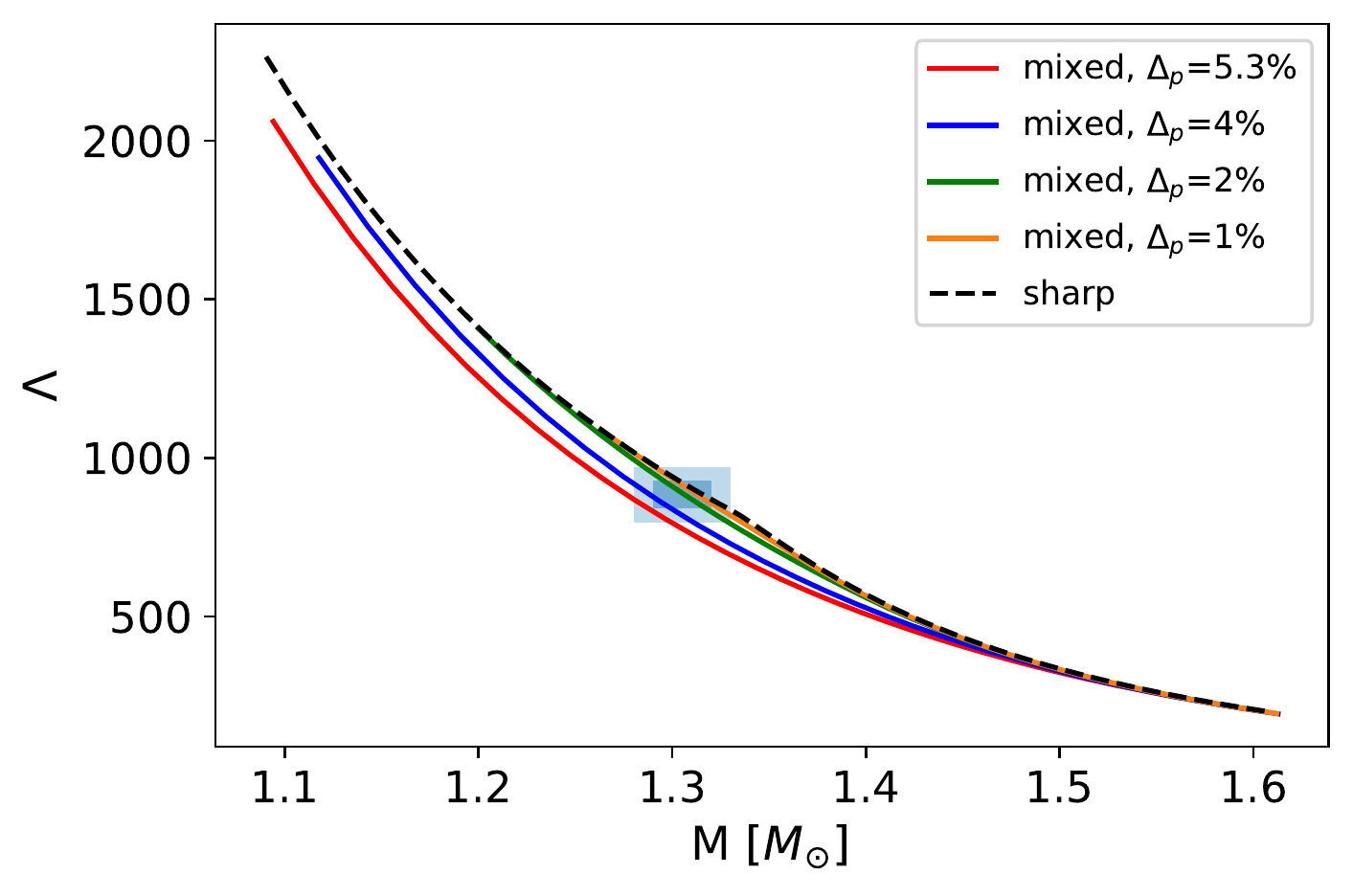}   
   \caption{Same as Fig. \ref{M_tidal_deformation_0805} but for weak phase transitions. The uncertainty boxes are centered at the same mass of Fig. \ref{M_R_0391} and its associated tidal deformability. Maximum fractional tidal differences here are much smaller than in the strong phase transition case.}\label{M_tidal_deformation_0391}
   \end{figure}

In Fig. \ref{M_R_071} we show a portion of the $M(R)$ relation for stars with $\eta=0.71$ (strong phase transition) and $\Delta_p$ from around $7\%$ (largest possible value as suggested from the surface tension analysis \citep{2018PhRvC..97d5802A}) to $1\%$, for masses and radii around the sharp phase transition.
The value $\Delta_p=6.7\%$ has been chosen for the parabolic construction because it coincides with the particularities of the polytropic EOS with a mixed state for $n_1=0.4\, {\rm fm}^{-3}$ in Fig.~\ref{fig:mixed4}. Figure \ref{M_tidal_deformation_0805} shows their associated tidal deformations as a function of the stars' masses. 
The boxes on the plots show the range of possibilities for the observables taking into account the minimum precision suggestions found in the previous section and
expectations of near-term and future detectors. 
Figures \ref{M_R_0391} and \ref{M_tidal_deformation_0391} show similar relations for $\eta=0.39$ (weak phase transition) by making use of the bag-like model with $c_s^2=1$ for the quark core as described before. In this case, the maximum $\Delta_p$ for which the parabolic construction works is $\sim 5.3\%$.

   %%%%%%%%%%%%%%%%%%%%%%%%%%%
   
From the above figures one can see that the benchmark accuracies for distinguishing a sharp-phase transition from a mixed state roughly agree for different constructions of the mixed state.
In the neighborhood of a strong phase transition, one has a mass range of stable stars with sharp phase transitions whose tidal deformabilities could differ up to around $25\%$ with respect to stars presenting mixed states (see Fig. \ref{M_tidal_deformation_0805}). However, differently from strong phase transitions, twin stars (see, e.g., \citep{2019PhRvD..99j3009M} and references therein) may not even exist if $\Delta_p$ is large enough. If this is not the case, then, depending on $\Delta_p$, twin stars (one of them being one-phased and the other one with a mixed state) might have any 
tidal deformability differences (they could be either zero--continuous--or not, as in the cases of $\Delta_p=3\%$ and $\Delta_p=1\%$, respectively, in Fig. \ref{M_tidal_deformation_0805}) around the mass marking the appearance of the quark phase (differently from the case of a given strong phase transition). In addition, the relative tidal deformability difference between a one-phase star and a hybrid star for a given mass could be as large as $50-60\%$, and that explains the large differences for strong phase transitions found in our analysis for polytropic EOSs in Sec. \ref{sect:polytropes_results}. 

Finally, similarly to Figs. \ref{fig:DeltaM_Deltap_pol}, \ref{fig:DeltaM_pres_pol} and \ref{fig:DeltaM_frac_pres_pol}, fractional changes to the maximum mass for different $\eta$ within the context of the parabolic construction for the mixed state are also up to ${\cal O } (0.1\%)$. Radius differences associated with maximum masses are also of ${\cal O}$(10 cm); relative changes of the radii are hence ${\cal O} (10^{-3}\%)$. All the above is expected based on the fact that for the maximum masses and associated radii the mixed state outcomes are almost indistinguishable from sharp-phase transition star under the same pressures and hence the particularities of an EOS construction are partially masqueraded.

\section{Discussion and conclusions} 
\label{sect:conclusions}

An interesting argument  in favor of a phase transition at intermediate densities (3-4 $\rho_{\rm sat}$, $\rho_{\rm sat}=2.7\times 10^{14}$g~cm$^{-3}$) may be derived from the precise measurement of the thickness of the neutron skin of $^{208}{\rm Pb}$ (see \citep{2021PhRvL.126q2503R} and references therein).  The skin thickness being larger than anticipated requires a stiffer EOS of neutron matter at $\rho\lesssim \rho_{\rm sat}$ \citep{2021PhRvL.126q2503R}. If smoothly continued to intermediate supranuclear densities, relevant for NSs of masses around $1.4 M_{\odot}$, this new EOS  is too stiff to be reconciled with measured values of $R$ and $\Lambda$, which require a softer EOS in this density region. This softening, followed by a stiffening at large densities---to allow for $2M_{\odot}$ NSs---is missing in the smooth EOSs based on   $^{208}{\rm Pb}$ skin measurements. According to \citep{2021PhRvL.126q2503R}, the tension between the two EOSs might indicate a phase transition at intermediate densities, relevant for NSs but irrelevant for $^{208}{\rm Pb}$.

Although expected under the theoretical point of view when the surface tension is below a critical value, there is not yet a direct observation for the presence of a mixed state in an NS. In this work we have tried to identify some observables that may evidence the mixed state. When it comes to radius and mass observations in general, it seems that only future missions with relative uncertainties smaller than $1\%-2\%$, may be relevant. Tidal deformabilities should also have high accuracies, with fractional uncertainties being smaller than $5\%-10\%$. (These accuracies also suggest the maximum systematic errors due to EOS modellings of hybrid stars.) However, for a range of masses close to the appearance of the quark phase in the case of strong phase transitions, tidal deformations associated with sharp interfaces and mixed states may differ more significantly. For instance, as evidenced by Fig. \ref{M_tidal_deformation_0805}, the relative change between a hybrid star with a stable quark core and another one with a mixed state and the same mass could be up to around $25\%$. 
Roughly, the above results agree for different constructions of a mixed state. 
All of this may give us hope to start probing the existence of the mixed state (or weakly constrain the high-density part of the EOS of a star) in the near future. In particular, the most promising region of the $M-R$ diagram for differentiating a sharp phase transition from a mixed state is around the phase transition mass, and in principle it could happen around and between the most commonly observed masses for NSs ($1.4M_{\odot}$ and $1.8M_{\odot}$ \citep{2018MNRAS.478.1377A}). However, statistical studies also suggest that the phase transition mass may be large, around $2M_{\odot}$ \citep{Annala:2019puf}. If this turns out to be the case, then tidal deformation measurements (if possible) and significant differences of them for less massive stars would hint that some of them may have mixed states in their interiors. (A mixed state should appear at a smaller pressure than the one marking the appearance of a quark phase in a star with a sharp phase transition. In addition, too massive stars may have tidal deformations too small to be measured even with advanced GW detectors.)

A promising way to disentangle sharp phase transitions from mixed states is with a large sample of observations/higher signal-to-noise ratios (SNRs) because it could put radius uncertainties down \citep{2021arXiv210812368C}.
We stress that the $1\%-2\%$ precision for radii (and $5\%-10\%$ for tidal deformations) would be meaningful when many observations (for different masses) are available. In this case, a particular sharp-phase transition EOS that could explain a few observations may not explain many. As a result, we would be able to better constrain some EOS aspects, in particular those with phase transitions. When this large number of observations will be available, one may also have an estimate for $\partial M/ \partial R$, which could also deliver some information about the level of softening of the EOS.

Characterizing a mixed state seems a much more complicated task than probing its existence. 
One of the reasons for different constructions leading to similar aspects is their sharing of key thermodynamic conditions, and also the fact that the mixed state's structure is only relevant for a limited range of chemical potentials. Importantly, in the range of masses where a sharp transition would differ the most from another one with a mixed state, different constructions for the mixed state are expected to lead to very small systematic uncertainties. In the example of Figs. \ref{M_R_071} and \ref{M_tidal_deformation_0805}, relative changes for the tidal deformations intrinsically associated with different models for the mixed state are much less than $1\%$. Fractional changes to the radius due to the mixed state modelling are much smaller, up to around $0.2\%$. This would suggest that third generation GW detectors and future electromagnetic missions may characterize some aspects to the mixed state in a rather model independent way.

At the macroscopic level relevant for NSs, the mixed state of dense matter is electrically neutral. However, on the microscopic scale, the space there is filled with structures of normal and exotic phases of opposite electric charges \cite{1996PhR...264..143G,PhysRevC.52.2250}. The Coulomb force is balanced out by the surface tension $\sigma$ between the two phases \cite{1996PhR...264..143G}, which is one of the key ingredients connecting  strong interactions of the matter constituents and the  phenomenology of a mixed state ($0<\sigma<\sigma_{\rm max}$ unknown) in a hybrid star. 
Coexisting substructures of exotic and normal phases, e.g., droplets, columns, plates and corresponding bubbly structures, are electrically charged and to minimize the energy in the mixed state, the equilibrium mixed state  has a periodic crystal ordering \cite{PhysRevC.52.2250}. The mixed state resists deformation via an elastic strain, which contributes to the matter stress tensor. In this way, the hydrostatic equilibrium  of an NS becomes the hydro-elastic one (e.g., \cite{2020arXiv201106361P}). A rough estimate of the maximum ellipticity of a solitary hybrid NS, supported by elastic strain of its mixed state was obtained by \cite{PhysRevLett.95.211101}. It will be of interest  to address this aspect of the mixed state and in particular its imprint on $\Lambda$, and we will do so in a follow up paper.

Our analysis suggests that differentiating a weak (sharp) phase transition from a mixed state will be much more observationally challenging. In case of a lack of observable differences between sharp and mixed EOS aspects might put upper limits on the density change from a hadronic to a quark phase (current multimessenger constraints are yet loose \citep{2021ApJ...913...27L,2021PhRvD.103f3026T,2021PhRvC.103c5802X}) and $\Delta_p$. These upper-limits could also be translated into limits to $\sigma$ given microscopic models. This would be relevant due to our current ignorance on this quantity. One may also roughly estimate the required SNR to differentiate between a weak phase transition and a mixed state. Figure \ref{M_tidal_deformation_0391} suggests that relative tidal deformation uncertainties should be at most of the order of $1-2\%$. Given that large SNRs scale inversely with uncertainties \citep{PhysRevD.98.124014}, for a GW170817-like event one would need an SNR $\approx 2000-3500$ to distinguish a phase transition with a mixed state from a weak (sharp) one. This is larger than the most optimistic expectations for the Einstein Telescope (see, e.g., \citep{PhysRevD.98.124014,2021arXiv210812368C} and references therein).

Other quantities worth exploring in order to single out aspects of the mixed state would be the moment of inertia of stars and their quadrupole moments. They would be relevant due to the prospect of near future measurements of the rate of advance of the periastron and ellipticities of stars, respectively. Indeed, it will be possible to measure the rate of the advance for some sources \citep{2020ApJ...901..155G}, and now we are closer to measuring mountains in NSs with GWs (see, e.g.,  \citep{2020ApJ...902L..21A,2021PhRvD.103f3019D,2021arXiv210909255T,2021arXiv211113106T} and references therein). In the vein of GWs, an elastic mixed state might be able to heighten mountains in the crust, and details thereof should be better understood. For third generation detectors, it would also be of interest to calculate the quasi-normal modes of a hybrid star with a mixed state because some modes might rise uniquely due to it in the range of hundreds to some kHz. In this case, the planned GW mission NEMO \citep{2020PASA...37...47A} might also be relevant. It is a 2.5-generation GW detector that will sacrifice sensitivity at low frequencies to obtain larger-than-current sensitivies in the high frequency band. As a result, it will be suitable for GW observations of the late inspiral and the post-merger phases of binary coalescences. Of particular interest is when one of the compact systems is an NS, because the detection of kHz GWs may unveil unique aspects pertaining to hybrid stars \citep{2019PhRvL.122f1102B,2020PhRvD.102l3023B}. In addition, when combined to Advanced LIGO and many detections are available, it may also be able to better constrain NS EOSs due to tidal deformation measurements (larger impact on the waveforms). The expected (fractional) radius precision is not far off from the one we have estimated to start evidencing the presence of a mixed state layer. Thus, mostly when many observations are present, NEMO may also have the potential of shedding light on some aspects of phase transitions in NSs.

Finally, we stress that precise predictions associated with a mixed state are not simple to be made when using masses, radii and tidal deformations of stars in general.\footnote{However, in particular, measurements of radii and tidal deformations near the phase transition mass might reveal the presence of mixed states in the most optimistic cases; see Fig. \ref{fig:mixed4}. If the softening of the EOS is identified, as we have assumed, one might in principle have an idea of the phase transition mass and observations in this mass range could be made.} Statistical results and high precision measurements for these observables are needed to assess the existence of a mixed state inside an NS. Estimating these accuracies, as we have done, is important as a first step to learn when one might run across limitations on the EOS recovery with $M, R$ and $\Lambda$ measurements and also to learn about the systematic uncertainties for EOS modelings with phase transitions. Smoking-gun effects for mixed states should be focused mostly on phenomena exclusively taking place in such a phase, for instance oscillation modes that could show up in the inspiral (pre-merger) and post-merger waveforms of gravitational waves of NSs and also on some electromagnetic measurements such as quasi-periodic oscillations and lightcurves. We leave such studies for future works.

\section{Summary}
\label{summary}
Hybrid stars with sharp phase transitions and mixed states may start being distinguished observationally in the most optimistic case either (i) using data from the GW detectors with tidal deformability  uncertainties smaller than $5-10\%$, suggesting that unless we witness rare nearby events with high SNRs, we need to rely in general on 2.5- and 3rd generation GW detectors, or (ii) using electromagnetic missions/GW detectors that could deliver radius (and masses) uncertainties smaller than 1-2$\%$. Measurements with higher uncertainties would lead to a limitation on the EOS recovery by means of NS masses, radii and tidal deformabilities. The above accuracies would also suggest the level of systematic uncertainties EOS models with phase transitions would have. The most promising cases concern strong phase transitions (large density jumps for the Maxwell construction) and mixed states with large $\Delta_p$ (the relative pressure change at the chemical potential related to the appearance of the quark phase for a sharp phase transition). Sharp weak phase transitions (smaller density jumps) seem more challenging to be observationally differentiated from stars presenting mixed states. 
In general, the mixed state would change negligibly the maximum masses of stars and their associated radii when compared to sharp phase transitions. The range of NS masses where changes between sharp phase transition and mixed state observables may be noticeable with near-term and future detectors is around the appearance of a quark phase, and particularities of the mixed state construction are expected to lead to very small systematic uncertainties. This suggests that constraints to the mixed state might be possible and rather EOS-free.

\begin{acknowledgments}
We thank Nikolaos Stergioulas and the anonymous referees for useful comments. The Authors gratefully acknowledge the financial support of the National Science Center Poland grants no. 2016/22/E/ST9/00037 and 2018/29/B/ST9/02013, and the Italian Istituto Nazionale di Fisica Nucleare (INFN), the French Centre National de la Recherche Scientifique (CNRS) and the Netherlands Organization for Scientific Research (NWO), for the construction and operation of the Virgo detector and the creation and support of the EGO consortium. 
\end{acknowledgments}

%\appendix

%\section{Appendixes}

\bibliography{bibliography}% Produces the bibliography via BibTeX.

\end{document}